\documentclass[final]{ar2e}
\usepackage{ulem}
\usepackage{graphicx}
\usepackage{subfigure}
\usepackage{color}
\usepackage{aas_macros}
\usepackage{ARAstroBib}
\usepackage{setspace}
\usepackage{url}
\begin{document}

\setlength\unitlength{0.0095\linewidth}  %%%do not remove

\newcommand{\Note}[1]{{bf #1}}
\newcommand{\ms}{{m\,s$^{-1}$}}
\newcommand{\kms}{{km\,s$^{-1}$}}
\newcommand{\acc}{{m\,s$^{-2}$}}
\newcommand{\bx}{{\bf x}}
\newcommand{\br}{{\bf r}}
\newcommand{\bu}{{\bf u}}
\newcommand{\bB}{{\bf B}}
\newcommand{\bk}{{\bf k}}
\newcommand{\id}{{\rm d}}
\newcommand{\bnhat}{{\bf \hat{n}}}
\newcommand{\zhat}{{\bf \hat{z}}}
\newcommand{\bxhat}{{\bf \hat{x}}}
\newcommand{\bkhat}{{\bf \hat{k}}}
\newcommand{\bxi}{\mbox{{\boldmath$\xi$}}}

\jname{Annu. Rev. Astron.  Astrophys.}
\jyear{2010}
\jvol{48}
\ARinfo{}

\title{Local Helioseismology: Three Dimensional Imaging of the Solar Interior}
\author{Laurent Gizon
%\footnote{Visiting Academic, Monash University, Clayton, Victoria, Australia}
\affiliation{
%Selbst\"{a}ntiger Nachwuchsgruppe `Helio- und Asterosismologie',
Max-Planck-Institut f\"ur Sonnensystemforschung,
Max-Planck-Stra{\ss}e 2, 37191 Katlenburg-Lindau, Germany; email: \url{gizon@mps.mpg.de}
}
Aaron C. Birch
\affiliation{NorthWest Research Associates, Colorado Research Associates Division, 3380 Mitchell Lane, Boulder, CO 80301, USA;  email: \url{aaronb@cora.nwra.com}}
Henk C. Spruit
\affiliation{Max-Planck-Institut f\"{u}r Astrophysik,
Karl-Schwarzschild-Stra{\ss}e 1, 85748 Garching, Germany; email: \url{henk@mpa-garching.mpg.de}}
}

\markboth{Gizon, Birch \& Spruit}{Local Helioseismology}


\begin{abstract}
The Sun supports a rich spectrum of internal waves that are continuously excited by turbulent convection. The GONG network and the MDI/SOHO space instrument provide an exceptional data base of spatially-resolved observations of solar oscillations, covering an entire sunspot cycle (11 years). Local helioseismology is a set of tools
for probing the solar interior in three dimensions using measurements of wave travel times and local mode frequencies. Local helioseismology has discovered (i)~near-surface vector flows associated with convection (ii)~$250$~\ms subsurface horizontal outflows around sunspots (iii)~$\sim 50$~\ms extended horizontal flows around active regions (converging near the surface
and diverging below), (iv)~the effect of the Coriolis force on convective flows and active region flows
(v)~the subsurface signature of the $15$~\ms poleward meridional flow,
(vi)~a $\pm 5$~\ms   time-varying depth-dependent component of the meridional circulation around the mean latitude of activity, and (vii)~magnetic activity on the far side of the Sun.
\end{abstract}

\maketitle

%The Sun supports a rich spectrum of internal waves that are continuously excited by turbulent convection. The GONG network and the MDI/SOHO space instrument provide an exceptional data base of spatially-resolved observations of solar oscillations, covering an entire sunspot cycle (11 years). Local helioseismology is a set of tools for probing the solar interior in three dimensions using measurements of wave travel times and local mode frequencies. Local helioseismology has discovered (i) near-surface vector flows associated with convection (ii) 250 m/s subsurface horizontal outflows around sunspots (iii) ~50 m/s extended horizontal flows around active regions (converging near the surface and diverging below), (iv) the effect of the Coriolis force on convective flows and active region flows (v) the subsurface signature of the 15 m/s poleward meridional flow, (vi) a +/-5 m/s time-varying depth-dependent component of the meridional circulation around the mean latitude of activity, and (vii) magnetic activity on the far side of the Sun.

\section{INTRODUCTION}
Helioseismology is the observation and interpretation of the
solar oscillations to
probe the solar interior. These oscillations, with periods around five minutes,
are due to the random superposition of acoustic waves and
surface-gravity waves and are excited by
turbulent convection in the upper layers of the Sun.
Solar oscillations were discovered by \citet{Leighton1962}
and interpreted by \citet{Ulrich1970} and \citet{Leibacher1971} as
internal acoustic waves trapped in spherical-shell cavities.
Wave motions are measured along the line of sight from the  Doppler shifts of
absorption lines in the solar spectrum.
A short review of solar oscillations is given in Section~\ref{observations}.

Helioseismology has produced a large number
of discoveries in solar, stellar, and fundamental physics.  It
provides the most precise tests of the theory of stellar
structure and evolution, for example it motivated a revision of the standard
model of particle physics to solve the solar neutrino problem.
Helioseismology also enables the study and discovery of effects that are not included
in standard solar models (standard models are spherically symmetric, non rotating, non
magnetic, and include a simplified treatment of convection).

One of the most exciting aspects of helioseismology is the search for
 the origin of the Sun's magnetic field,
one of the most important unsolved problems in solar physics.
The eleven-year solar magnetic cycle is thought
to be due to a field-amplification (dynamo-) process \citep[cf.][]{Charbonneau2005,Rempel2008JPCS}, whereby a toroidal magnetic field component (in the azimuthal direction with respect to the rotation axis) is built up by stretching of the field lines by the Sun's differential rotation. In a second step the toroidal field is partially converted into a poloidal field component, which 'closes' the dynamo cycle.
Models for this second step (`$\alpha$-effect') differ significantly. In most models, it is attributed to the effect of convection on the magnetic field (convective dynamos). In an older model more closely connected with observations, convective flows play no role in this step; it is instead due to the instability of the toroidal field itself. The instability causes loops of magnetic field to rise to the solar surface, and appear as the observed magnetic (sunspot) activity.  Resolution of this conflict between the models is key for progress towards a theory of stellar magnetic fields which has real predictive power. Helioseismology holds the promise of providing new observational constraints on cycle-related structures below the surface
\citep[e.g.][]{Kosovichev2008}.

Traditionally, helioseismology methods have been classified into two groups: global helioseismology and
local helioseismology. Global helioseismology consists of
measuring the frequencies of the modes of oscillation and searching for
seismic solar models whose oscillation frequencies match the observed ones
\citep[see][for a review of techniques and results]{JCD2002}.
Global helioseismology is two dimensional and is used to infer solar properties as functions of radius and unsigned latitude.
A major achievement of global helioseismology is
the inference of the angular velocity
in the solar interior \citep[e.g.][]{Schou1998,Thompson2003}.
The differentially-rotating convection zone
and the rigidly-rotating radiative interior
are separated by a transition region at $0.69 R_\odot$, the tachocline,
which may be the seat of the solar dynamo.

Unlike standard global mode helioseismology, local helioseismology is capable of probing the solar interior
in three dimensions. This is important for the  study of solar activity, which is seen on the surface as
localized patches of magnetic field, e.g. active regions, sunspots, and plage regions.
Local helioseismology can potentially be used to infer vector flows, thermal and structural
inhomogeneities, and even the magnetic field itself. Local helioseismology
has been reviewed by e.g. \citet{Kosovichev1997SCORE},
\citet{Braun2000}, \citet{Kosovichev2000}, \citet{JCD2002},
\citet{Kosovichev2002},  %\citet{Birch2006}, \citet{Gizon2006},
\citet{Komm2006ASR}, \citet{Gizon2007AN}, \citet{Birch2008}, and
\citet{Thompson2008SOL}.
The most comprehensive review is provided by \citet{Gizon2005}.

Local helioseismology encompasses various methods of data analysis
(Section~\ref{sec.methods}).
One method of local helioseismology, ring diagram analysis, is a
relatively straight-forward extension of global helioseismology. It
consists of measuring local frequencies of oscillation by analyzing
small patches on the Sun. Ring-diagram analysis is computationally efficient
and has produced important results, such as
maps of flow patterns in the Sun.

Other methods of local helioseismology, like time-distance
helioseismology and helioseismic holography, are based on the computation of the
cross-covariance between the oscillation signal measured at two points on the surface.
The basic principle is to retrieve information about the solar interior from the time it takes for
solar waves to travel between any two surface locations through the solar
interior. The cross-covariance function is directly related to the
Green's function and thus carries essential information.

Like in global helioseismology, an inverse problem must
be solved in order to retrieve subsurface solar properties
from local helioseismic measurements (Section~\ref{sec.forward_inverse}).
In many cases it is acceptable to assume that the Sun is weakly heterogeneous in the horizontal directions:
the inverse problem becomes a linear inverse problem and can be
solved with standard techniques.
However this is not always possible, especially in the presence of strong
magnetic fields, e.g. in a sunspot.

Time-distance helioseismology aims at inferring subsurface properties at the
best spatial and temporal resolution possible. A spatial resolution as small
as a few Mm can be achieved near the surface. This limit is
intimately linked to the smallest available horizontal wavelength of the solar oscillations
(high frequency surface gravity waves).  Detailed 3D maps of vector flows
in the upper convection zone have provided new insights into the structure,
evolution and organization of magnetic active regions and convective flows.
The most easily detectable flow pattern is supergranulation, an
intermediate-scale of convection (Section~\ref{sec.superg}).

A particularly challenging aspect of local helioseismology is sunspot
helioseismology (Section~\ref{sec.sunspots}).  Sunspots are regions of intense (kilogauss) magnetic field and low gas pressure and density.
In spite of an abundance of telling clues from observations of the solar surface, theories about their formation, subsurface structure, thermal
properties, and deep magnetic field topology are still controversial.
The nature of solar waves is very significantly altered as they propagate through a sunspot and convert into magneto-acoustic-gravity (MAG) waves. Numerical modeling of wave propagation through model sunspots is currently being developed by several groups. These simulations will be key to interpret the solar oscillations in the vicinity of sunspots.
Realistic numerical simulations also promise to be an important diagnostic tool for sunspot structure.
The main question---what keeps a sunspot together as a clearly delineated entity?---may not ultimately be answerable by helioseismology, since key elements of the answer may well lie in a region near the base of the convection zone, where helioseismological tools may not have enough sensitivity to detect a sunspot-related signal. They may perhaps be sufficient, however, to challenge models that propose the origin of sunspots to be in the surface layers.

Among the most interesting results of local helioseismology is
the detection of the subsurface meridional flow
(Section~\ref{sec.global_scales}). The meridional flow does not
affect global mode frequencies (to first order) and thus has only
 been measured in the solar interior with local helioseismology.
 The meridional flow plays an important role in 'flux transport' theories, according to which the latitudinal transport of magnetic flux at the base of the convection zone determines the period of the solar cycle.

In yet another remarkable application, local helioseismology can be used to
construct maps of active regions on the far side of the Sun (Section~\ref{sec.farside}). In farside imaging, the Sun as whole is used
as an acoustic lens focussing waves at a point on the invisible hemisphere.
Maps of the farside are potentially important to
predict space weather and provide advance warning for coronal mass ejections and solar flares
(violent and sudden release of
energy associated with the reconfiguration of the magnetic
field in the atmosphere above an active region).
Flares can excite acoustic waves to measurable levels, which can in turn
tell us about the physics of flares (Section~\ref{sec.flares}).

\begin{figure}[t!]
\includegraphics[width=13.cm]{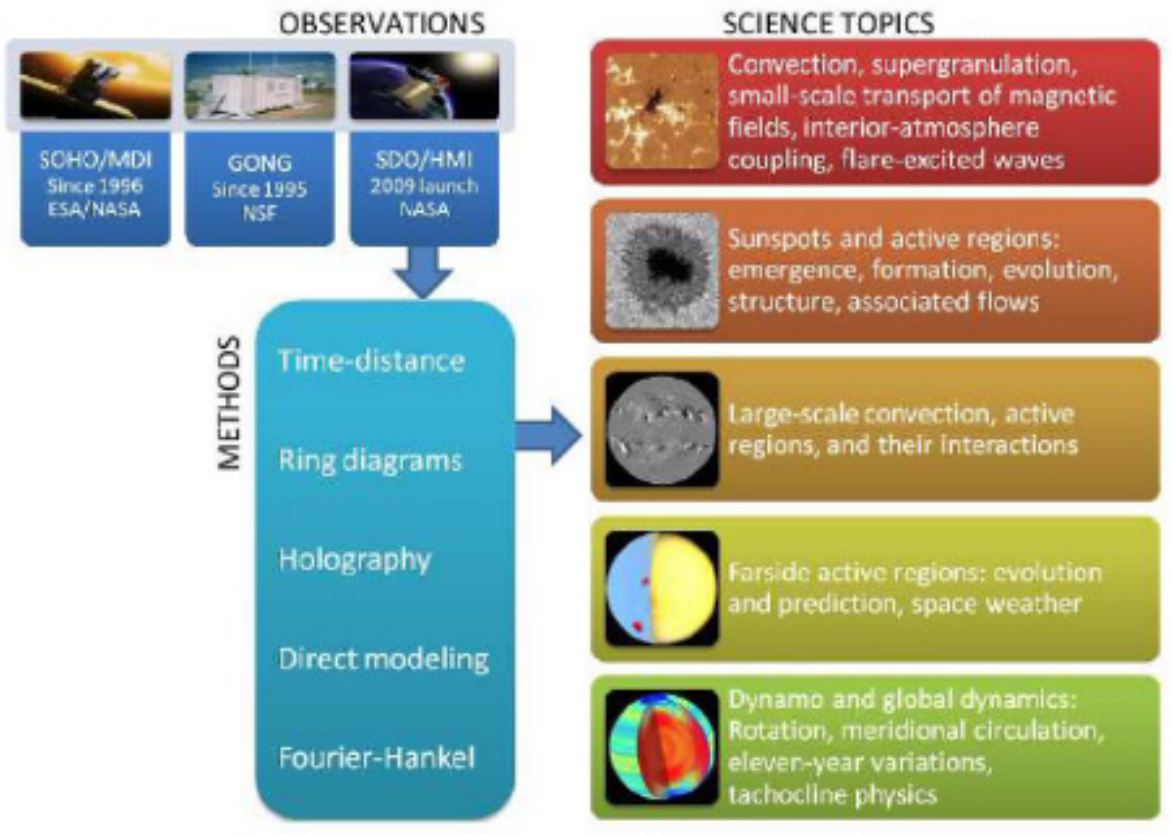}
\caption{
Overview of local helioseismology: Observational data, methods of analysis, and scientific applications.
%{\bf NOTE TO EDITOR: original pictures are named Gizon\_f1\_*.*.}
}
\label{fig.overview}
\end{figure}

These examples illustrate the many facets of the science possible with local helioseismology,
as summarized in {\bf Figure~\ref{fig.overview}}.
In all these cases a taste of the possibilities has been provided,
but improved observations (Section~\ref{sec.future})  and
further developments in the techniques of analysis and interpretation
are required to realize the full potential of local helioseismology.

%%%%%%%%%%%%%%%%%%%%

\section{SOLAR OSCILLATIONS}
\label{observations}
\subsection{Observations}
In most cases, local helioseismology uses time-series of Dopplergrams as input data.
A Dopplergram is a digitized image of the line-of-sight velocity of the solar surface (photosphere or chromosphere) deduced from the Doppler shifts of a Fraunhofer absorption line
\citep[e.g.,][]{Scherrer1995}. Solar oscillations have a higher signal-to-noise ratio in
Doppler velocity than in intensity, especially at low frequencies.

There are two major data sets available for local helioseismology.
The first one is provided by the Global Oscillation Network Group
\citep[GONG,][]{Harvey1996}
headquartered in Tucson, Arizona, which operates a global network of six
stations around the world. The sites are distributed in longitude in order
to observe continuously:  Big Bear (California), Mauna Loa (Hawaii),
    Learmonth (Australia),  Udaipur (India), El Teide (Canary Islands), and
    Cerro Tololo  (Chile).
The cadence of the observations is
one minute to avoid temporal aliasing.
Each GONG instrument is a phase-shift interferometer that measures
the phase of the Fourier transform of the
solar spectrum around the Ni I absorption line at 6768~\AA,
 interpreted as a Doppler shift \citep{Harvey1995}.
While the original cameras had an image size of 256$\times$256 pixels,
 full-disk Dopplergrams have been recorded with 1024$\times$1024 CDD cameras
since 2001, hence providing
a good spatial resolution ($5$ arcsec) for local helioseismology.
The GONG instruments also acquire intensity images and line-of-sight
magnetograms. The effective duty cycle of the merged observations is over 90\%.

\begin{figure}[t!]
\includegraphics[width=\linewidth]{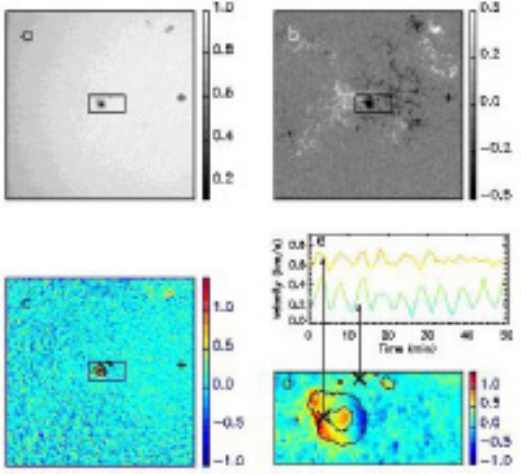}
\caption{SOHO/MDI observations on 22 January 2008 at 17:36:00 remapped using Postel's azimuthal equidistant projection with a map scale of $0.12$~deg/pixel or $1.46$~Mm/pixel. The sunspot in Active Region NOAA 9787 is at the center of projection. ({\it a}) 512$\times$512 pixel subfield of the continuum   intensity, normalized to unity at disk center (plus sign). The box around the sunspot has size 147 Mm $\times$ 73 Mm. ({\it b}) Line-of-sight component of the magnetic field in kG (truncated gray scale). ({\it c}) Line-of-sight Doppler velocity in \kms. Supergranulation is visible toward the edges of the frame. ({\it d}) Doppler velocity in the sunspot box. The black contours give the outer edges of the umbra and penumbra of the sunspot. The center-to-disk component of the Evershed outflow is visible in the penumbra. ({\it e}) Doppler velocity as a function of time at the two locations denoted by the crosses in panel {\it d}. The five minute period of the solar oscillations is evident. The oscillations have reduced amplitudes in the sunspot.}
\label{fig.observables}
\end{figure}

The other main data set is provided by the Michelson Doppler Imager \citep[MDI,][]{Scherrer1995}
on-board the ESA/NASA Solar and Heliospheric
Observatory (SOHO), which was launched in December 1995. SOHO is in a halo
orbit around the Sun-Earth L1 Lagrange point. Observations from SOHO are not only
 continuous, but benefit from perfect seeing
and from a slowly varying spacecraft-to-Sun velocity.
The MDI filter system relies on two tunable Michelson interferometers in order to
measure intensity in five very narrow filters (94~m\AA)
in the wings and core of the Ni~6768 line.
The Doppler velocity is obtained by taking the difference between filtergrams
on each side of the absorption line.
The MDI observables are the line depth and continuum, line-of-sight Doppler
velocity, and line-of-sight magnetic field.
The temporal cadence is one minute and the CCD camera has 1024$\times$1024
pixels. It can operate in two different modes: a full-disk mode ($2$ arcsec
pixel) or a high-resolution mode ($0.6$~arcsec pixel).
{\bf Figure~\ref{fig.observables}} shows example full-disk SOHO/MDI observables
and {\bf Supplemental Movie 1} shows a time series of full-disk Dopplergrams.
In high-resolution mode, the resolution is better
by a factor of three but the field of view is reduced.
Because of limited telemetry, the full-disk Dopplergrams have only
been transmitted at full cadence for about two to three months each year since
1996, while the high-resolution Dopplergrams are reserved for targeted
campaigns of observation.
The rest of the time, the Dopplergrams are spatially filtered onboard
and converted into lower-resolution $256\times256$ images,
in order to save  telemetry ('medium-degree' data).

GONG and MDI are complemented by other data sets, e.g. from the Taiwan
Oscillation Network \citep[TON,][]{Chou1995}, from campaigns of observations
at the South Pole with the Magneto-Optical filter at Two Heights  \citep[MOTH,][]{Finsterle2004SOL}, and from the Hinode satellite \citep[e.g.][]{MitraKraev2008}.

In many applications of local helioseismology, the standard procedure consists
of choosing a relatively small region of the Sun and following (or `tracking') it
in a frame that is co-rotating with the Sun.
This gives a time series of Doppler images that are centered on the
region of interest, like a magnetic active region.
In this process each individual image is mapped onto a common spatial grid.

For local studies it is often convenient to neglect the curvature of the solar
surface and work in plane-parallel geometry. With this simplification,
it is natural to study the oscillations in three-dimensional Fourier space.
The oscillation signal, denoted by $\phi(\br,t)$, where $\br=(x,y)$ is the
horizontal position vector and $t$ is time, is decomposed into harmonic
components
\begin{equation}
\phi(\bk, \omega) = \int_A \id^2\br \int_0^T \id t  \; \phi(\br,t) \; e^{- i \bk\cdot\br + i \omega t} ,
  \label{eq.fft}
\end{equation}
where $A$ is the area of study, $T$ is the total observation time, the vector $\bk=(k_x, k_y)$ is  the horizontal wavevector, and $\omega$ is the
angular frequency. The horizontal wavenumber is
$k=\|\bk\|$. By convention, the $x$ coordinate is positive in
the direction of rotation (prograde) and the $y$ coordinate points north.
The power spectrum of solar oscillations is defined as
\begin{equation}
P(\bk, \omega) =  |\phi(\bk, \omega)|^2 .
\label{eq.power}
\end{equation}
We note that the spatial Fourier transform should be replaced by a spherical
harmonic transform when curvature effects cannot be ignored, as in global helioseismology.

\begin{figure}[t!]
%\begin{picture}(100,70)
%\put(6,6){\includegraphics[width=12.cm, trim=1.7cm 1.1cm 0 0, clip]{Gizon_f3.eps}}
%\put(50,1){$k_x R_\odot$}
%\put(0,33){\rotatebox{90}{$\omega/2\pi$ (mHz)}}
%\end{picture}
\includegraphics[width=\linewidth]{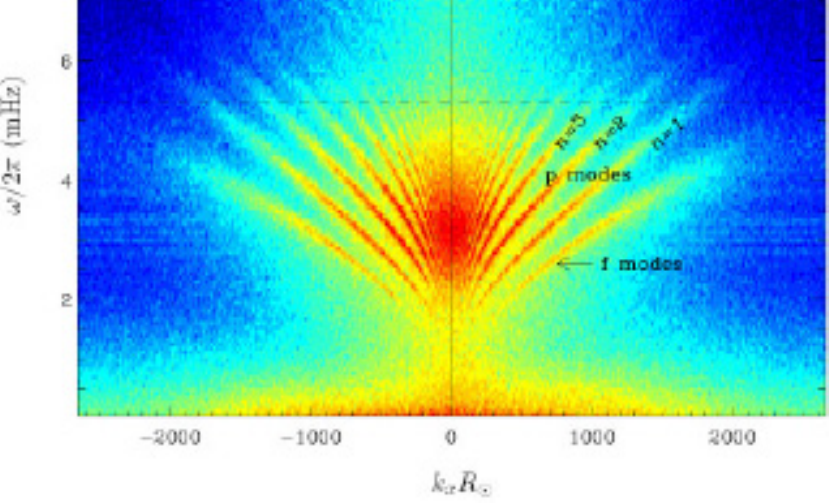}
\caption{
Cut at $k_y=0$ through an average power spectrum of MDI/SOHO high-resolution Doppler velocity data
as a function of frequency and $k_x R_\odot$. The horizontal dashed line shows the acoustic cutoff frequency. In order to reduce random noise, an average was carried out over eight individual power spectra, each of duration $T=4$~hr and covering an apodized region of area $A\sim(500 \; \mbox{Mm})^2$. Since $k_y=0$, only waves traveling in the east or west directions are showed. The power below $\sim 1.5$~mHz is due to solar convection, granulation and supergranulation.
}
\label{fig.power}
\end{figure}

An example power spectrum of solar oscillations is shown in
{\bf Figure~\ref{fig.power}}. Power is distributed along well-defined
discrete ridges in wavenumber-frequency space
and peaks around 3~mHz.
The first ridge at low frequencies shows the ``fundamental'' (f) modes.
These are surface gravity waves with exponential eigenfunctions
and a dispersion relation $\omega^2 = g k$, where $g=274$~\acc
is the
acceleration of gravity at the solar surface;
they are similar to waves at the surface of a deep ocean.
All other ridges correspond to pressure (p) modes, i.e. acoustic waves
modified by gravity.
The existence of discrete ridges, $\omega=\omega_n(k)$ with $n>0$, reflects the fact that p modes
are trapped in the vertical direction.
At fixed wavenumber, the peaks of power are labeled p$_0$, p$_1$, p$_2$, etc.\ with increasing frequency.
A mode p$_n$ is such that the number of radial nodes of the mode displacement is $n$ (the radial order).
By convention, the f modes are labeled with $n=0$.
All ridges have reduced power above $5.3$~mHz, which is the cut-off frequency above which
waves are not reflected back into the Sun but escape into the atmosphere.
The frequency width of a ridge is inversely proportional to the mode lifetime.
A recent description of the mode parameters, including mode lifetimes, is provided by
\citet{Korzennik2004}.

\subsection{Modes}
In order to better understand the diagnostic capability of each mode, it is
 useful to consider simple solar models. For our purpose, a
simple solar model is a reference standard solar model,
which only depends on height (or radius), such as Model S \citep{JCD1996}.
In plane-parallel models that are isotropic and
translation invariant in the horizontal directions, the normal
modes of the oscillations of the model vary horizontally as $\exp(i \bk\cdot\br)$.
For the case of p and f modes, it is convenient to introduce the mode
eigenfunctions
$U_n(z;k)$ and $V_n(z; k)$ such that the
complex displacement eigenfunction of the mode characterized by
radial order $n$ and horizontal wavevector $\bk$ is
\begin{equation}
\bxi_n(\br,z; \bk) = \left[ \zhat U_n(z; k) + i \bkhat V_n(z; k)  \right]
e^{i \bk\cdot\br} ,
\end{equation}
where $z$ is height,  $\zhat$ is the unit vector pointing upwards,
and $\bkhat$ is the horizontal  unit vector pointing in the direction of
$\bk$.  Zero height corresponds to the photosphere ($-z$ is depth).
The representation
of the displacement eigenfunctions in terms of only the two functions $U$ and
$V$ is possible as neither the f nor the p modes have horizontal motions that
are perpendicular to $\bk$. As a result of the assumed
isotropy, the functions $U$ and $V$ do not depend on the direction of
$\bk$ and the mode frequencies $\omega_n(k)$ only depend on
wavenumber. The time evolution of the mode $(n,\bk)$ is given by $\exp[-i \omega_n(k) t]$.
In models that include attenuation, the frequencies are complex, while they are real for the case of adiabatic oscillations and standard boundary conditions.

\begin{figure}[t!]
\includegraphics[width=13.cm]{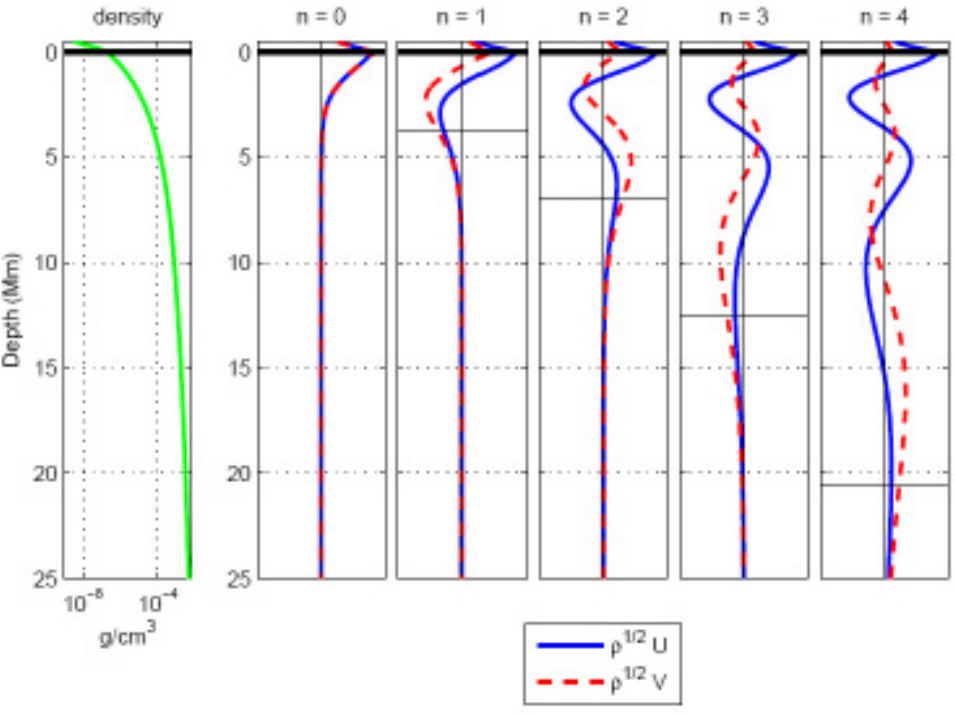}
\caption{
Density profile from Model~S (left panel, green line) and mode eigenfunctions $U$ and $V$ for the radials order $n=0$\,--\,$4$ (other panels) at the frequency $3.5$~mHz.  The lower turning points of the modes $n=1$\,--\,$4$ are shown as thin horizontal black lines. At fixed frequency, the horizontal phase speed $\omega / k$ increases with increasing radial order $n$, and therefore lower turning points increase with increasing $n$ as well. The functions $U$ and $V$ have been scaled with $\rho^{1/2}$ as the kinetic energy density is proportional to $\rho ( U^2 + V^2 )$.
}
\label{fig.eigenfunctions}
\end{figure}

{\bf Figure~\ref{fig.eigenfunctions}} shows the horizontal and vertical
eigenfunctions corresponding to the first five radial orders at a frequency
of $3.5$~mHz.  The solar model for this case is a plane-parallel version of
Model~S.  The eigenfunctions are scaled by $\rho^{1/2}$,
where $\rho$ is the density (left-most panel).
This scaling is used as we are interested in the kinetic
energy density of the modes, $\rho(U^2 +V^2)$, which is a physically
relevant quantity. For the f mode the horizontal and
vertical displacement eigenfunctions are equal.  For the acoustic
modes ($n>0$), the lower turning point, $z_{\rm t}$,  is the height at which
the sound speed is equal to the horizontal phase speed of the mode: $c(z_{\rm
  t}) = \omega / k $ (neglecting the buoyancy frequency and the acoustic
cutoff frequency, both of which are very small below a few~Mm beneath the
photosphere).
Thus all the modes with a similar horizontal phase speed (a straight line
through the origin in {\bf Figure~\ref{fig.power}}) have a similar lower turning point and probe essentially the
same layers of the Sun.

\section{LOCAL HELIOSEISMOLOGY}
\label{sec.methods}

In this section we give an overview of the various methods of local helioseismology. For an in-depth description of each method see e.g.~\citet{Gizon2005} and references therein.

\subsection{Ring Diagram Analysis}
The first operation in ring diagram analysis is to cover some fraction of the visible solar disk with patches (overlapping or not) with circular areas with diameters in the range $2^\circ$\,--\,$30^\circ$. Each patch is tracked in longitude with a velocity close to the solar surface rotation velocity to produce a time series of helioseismic observations (Dopplergrams or intensity images). For each patch a three-dimensional local power spectrum of the solar oscillation, $P$, is computed according to Equations~\ref{eq.fft} and~\ref{eq.power}.
The local power spectra reflect the local physical conditions in the solar interior, such as wave speed and horizontal flows \citep{Hill1988}. For example, a constant horizontal flow $\bu$ will introduce a Doppler shift of the power spectrum:
\begin{equation}
P(\bk, \omega) = P_0(\bk, \omega-\bk\cdot\bu) ,
\label{eq.pring}
\end{equation}
where $P_0$ is the power spectrum in the absence of a flow.
This description is highly simplified as flows in the Sun do vary with horizontal position and depth.
A change in the structure of the solar interior produces a change in the dispersion relation that does  not depend on the direction of $\bk$, and thus a change in the power spectrum that is also independent of the direction of $\bk$.

\begin{figure}[t!]
\includegraphics[width=\linewidth]{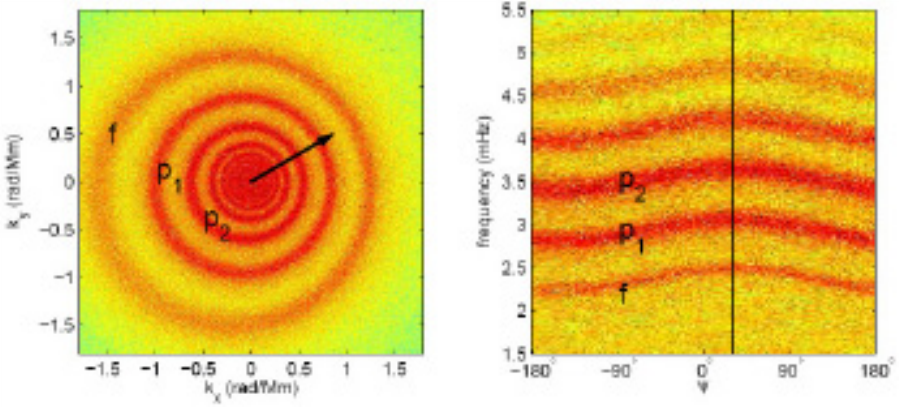}
\caption{
Slices through a model local power spectrum at constant frequency
$\omega/2\pi=3.1$~mHz ({\it left}, ring diagram) and at constant wavenumber $k=0.8$~rad/Mm ({\it right})
for the case of a depth-independent horizontal flow $\bu$ with an amplitude of
1~km/s and in the direction $\psi_0$ that is
thirty degrees north of the prograde direction.  In the left panel,
the black arrow shows the
direction of the flow.  The different rings correspond to different
radial orders; the outermost
ring is the f mode.  The rings with large $k$ are more strongly
influenced by the flow than those with small $k$ (see text).
In the right panel, the ridge frequencies show a sinusoidal variation
with $\psi$ and reach
their maxima when $\psi=\psi_0$ (shown by the vertical black line).
The frequency variation with $\psi$ is the same for all of the ridges
as as the flow is independent of depth.
}
\label{fig.frankring}
\end{figure}

There are two ways to study local power spectra.
The first approach is to consider cuts at constant frequency, $\omega$. In $(k_x,k_y)$ space, wave power is concentrated in rings, each corresponding to a different radial order $n$ \citep{Hill1988}. A ring diagram is shown in {\bf Figure~\ref{fig.frankring}}.
When there is no flow, the radius of each ring is the wavenumber $k$ at which $\omega_n(k)=\omega$ and is sensitive to the local dispersion relation. Thus the ring radius is related to the local wave speed under the patch.
As shown above, a flow will affect the local power spectrum. Linearizing Equation~\ref{eq.pring} for small $\bk\cdot\bu$, we find that the change in the ring position is  $\delta \bk = - (k / v_{\rm g})  \bu$, where $v_{\rm g} = \partial_k \omega_n$ is the group speed.
%As  \delta P = k u \partial_\omega P  = k u vg \partial_k P  =  k u vg  delta p / delta k
Hence the flow amplitude and direction can be estimated from the distortion and orientation
of the distorted rings. Note that the distortion of a ring depends on radial order in the case of a depth-varying flow.

The second approach consists of considering cuts at constant wavenumber, $k$, through the power spectra \citep{schou1998ring}.
The local power spectra are then studied in $(\psi,\omega)$ space, where $\psi$ is the azimuth of the wave vector measured from the prograde direction, $\bxhat$. The modes appear as bands of power around the resonant frequencies $\omega_n(k)$.
According to Equation~\ref{eq.pring}, a constant horizontal flow will Doppler shift the mode frequencies by
$\delta \omega = \bk\cdot\bu = k u_x \cos\psi + k u_y \sin \psi$.
As a result, $\bu$ can be estimated from the frequency shifts at each $k$ and radial order $n$.
A change in the wave speed will simply manifest itself as a change in the wave frequencies that is independent of $\psi$ and can be disentangled from the effect of a horizontal flow.

Both approaches rely on fitting a parametric model of the power spectrum to the observed local power spectra. Several functional forms have been proposed to fit the observations. The most important fitted parameters are the mode frequencies, $\overline{\omega}(n,k)$, and the two flow parameters, $\overline{u}_x(n,k)$ and $\overline{u}_y(n,k)$.
There is one set of parameters for each wavenumber and radial order. Details about the fitting procedures are given by, e.g.,  \citet{Basu1999} for the first approach and, e.g., \citet{Haber2000} for the second approach.

The fitted parameters are sensitive to the conditions in the solar interior, with a depth sensitivity that depends on the eigenfunction of the mode \citep[e.g.][]{JCD2002}. For example, the depth sensitivity to a horizontal flow is approximately given by the kinetic energy density of the mode \citep[e.g.][]{Birch2007APJ}.   The larger the horizontal phase speed of the mode, the deeper the sensitivity. Thus large patches give access to deeper regions in the Sun than small patches.
The differences between the fitted mode frequencies, $\overline{\omega}$,  and the mode frequencies calculated from a standard solar model, $\omega_n(k)$, are used in one dimensional (depth) inversions to infer structural conditions under each patch \citep[e.g.][]{Basu2004}. Two independent structural quantities can be inverted at a time, e.g., sound speed and density, from which other quantities can be inferred, such as the first adiabatic exponent. Similarly, the depth dependence of the horizontal flows, $u_x$ and $u_y$, can be inferred from a set of fitted parameters  $\overline{u}_x$ and $\overline{u}_y$ \citep{Hill1989}. Typically, ring analysis is used to probe the top 30~Mm of the convection zone, with a maximum horizontal resolution of about $2^\circ$ near the surface. Three-dimensional maps can then be obtained by combining neighboring patches.

\subsection{The Cross-Covariance Function}
\label{sec.xc}
\begin{figure}[t!]
\includegraphics[width=7.3cm]{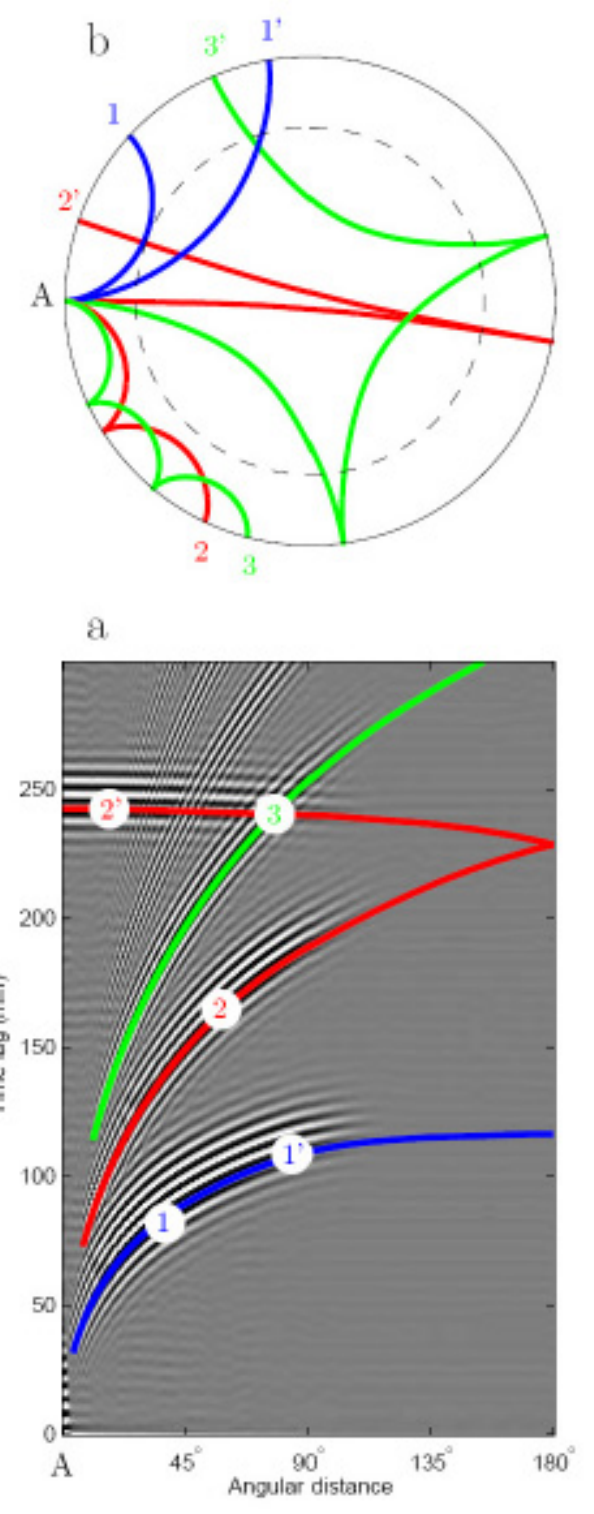}
\caption{({\it a}) Measured cross-covariance function for MDI medium-degree data as a function of separation distance and time-lag \citep{Kosovichev2000}. Positive values are white, negative values black. The observation duration is $T=144$~days starting in May 1996. ({\it b}) Example ray paths for acoustic wave packets. In both panels~({\it a}) and~({\it b}) the blue lines correspond to single skip ray paths, the red lines are for two skip paths, and the green lines are for three-skip paths.}
\label{fig.ray_paths}
\end{figure}

Time-distance helioseismology is based on the measurement of the
cross-covariance between the Doppler signals at
two points $\br_1$  and $\br_2$ on the solar surface,
\begin{equation}
C(\br_1, \br_2, t) = \int_0^T \id t'  \; \phi(\br_1, t')\phi(\br_2, t'+t) \; ,
\end{equation}
where $t$ is the correlation time-lag.
{\bfseries Figure~\ref{fig.ray_paths}}{\boldmath $a$} shows a cross-covariance function
measured from 144~days of MDI medium-degree data.  The cross-covariance has been
averaged over many pairs of points $( \br_2, \br_1)$ and is presented as
a function of the heliocentric angle between these two points.
This diagram is known as the "time-distance diagram."
The cross-covariance is essentially a phase coherent average of the random
oscillations \citep[][and {\bf Supplemental Movie 2}]{Bogdan1997}. It is a solar seismogram:  it provides a way to measure wave travel times between two surface locations.

A particular wave packet (consisting of set of modes with similar
phase speeds) is preferentially selected for each
travel distance.   Many applications involve much less temporal and
spatial averaging than was used in this figure.
Typical cross-covariances are therefore much noisier than the example shown
here.
The deeper meaning of the cross-covariance was elucidated only recently
in terms of Green's functions (see {\bf Side Bar}).

%\begin{figure}[h!]
%\begin{picture}(103,155)
%\put(0,1){\includegraphics[width=8.cm]{Gizon_f6_b.eps}}
%\put(7.8,98){\includegraphics[width=7.cm, angle=90.]{Gizon_f6_a.eps}}
%\put(7.3,2){\large A}
%\put(5.2,124.2){\large A}
%\color{blue}
%\put(14.5,144){1}
%\color{white} \put(19.,30){\circle*{5}} \color{blue}
%\put(18.,29){1}
%\put(30.2,152){1'}
%\color{white} \put(32.,37){\circle*{5}} \color{blue}
%\put(31.,36){1'}
%\color{red}
%\put(8,134){2'}
%\color{white} \put(13.3,72.5){\circle*{5}} \color{red}
%\put(12.3,71.5){2'}
%\put(22,98){2}
%\color{white} \put(24.8,52){\circle*{5}} \color{red}
%\put(23.8,51){2}
%\color{green}
%\put(23,150){3'}
%\put(27,96.5){3}
%\color{white} \put(30.3,72){\circle*{5}} \color{green}
%\put(29.3,71){3}
%\color{black}
%\put(11,150){\LARGE b}
%\color{white}
%\put(11,85){\LARGE a}
%\color{black}
%\end{picture}
%\caption{\tiny ({\it a}) Measured cross-covariance function for MDI medium-degree data as a function of separation distance and time-lag \citep{Kosovichev2000}. Positive values are white, negative values black. The observation duration is $T=144$~days starting in May 1996. ({\it b}) Example ray paths for acoustic wave packets. In both panels~({\it a}) and~({\it b}) the blue lines correspond to single skip ray paths, the red lines are for two skip paths, and the green lines are for three-skip paths.}
%\label{fig.ray_paths}
%\end{figure}

An important tool for visualizing wave
propagation in the Sun is the ray approximation \citep[e.g.][]{Kosovichev1997SCORE}.
In this approximation, the wavelength is treated as if it were much smaller than the
length scales associated with the variations in the background solar model.
The ray paths describe the propagation of wave energy and are
analogous to the rays in geometrical optics.
For discussions of  the range of validity of the ray approximation see e.g.~\citet{Hung2000} and \citet{Birch2001}.

{\bf Figure~\ref{fig.ray_paths}}{\boldmath $b$} shows some example ray paths computed
from Model~S. In this figure, the rays all begin from the same point at the
solar surface. Downward propagating rays are refracted by the increase of the sound speed
with depth until they reach their lower turning where the horizontal phase speed matches the sound speed.  At frequencies below about $5.3$~mHz,  upwards propagating rays are reflected from the solar surface by the sharp rise in the acoustic cutoff frequency.  At higher frequencies, the waves escape into the solar atmosphere.

The main features in the time-distance diagram
({\bf Figure~\ref{fig.ray_paths}}{\boldmath $a$})
are the ridges which correspond to different paths
that wave energy takes between pairs of observation points.
For example, the blue line corresponds to "first-bounce" arrivals (i.e., waves
that visit their lower turning points once between the two observation
points). The fine structure of the ridges in the time-difference diagram
reflects the band-limited nature of the power spectrum. The majority of the
wave power is near 3~mHz and as a result the cross-covariance shows fine structure that has a period of about 5~min.
The other ridges seen in the time-distance diagram correspond to multiple
bounces.   One particularly important ray path is the third bounce ray path
(green $3'$) that travels to the farside of the Sun before returning to the
visible disk, this ray path plays a central role in farside imaging (Section~\ref{sec.farside}).
As can be seen in the example ray paths in {\bf Figure~\ref{fig.ray_paths}}{\boldmath $b$}, the depth of the lower turning point increases with the distance the ray travels in a single skip.

We note that  the cross-covariance is directly related to the local power spectrum in the case when the medium is assumed to be horizontally invariant over this local area. In this case, the cross-covariance is simply given by the inverse Fourier transform of the local power spectrum \citep{Gizon2002}:
\begin{equation}
C(\br_1,\br_2,t) = {\rm const.} \; \int \id^2\bk \int \id\omega \;  P(\bk, \omega) e^{i \bk\cdot(\br_2-\br_1) - i \omega t} .
\label{eq.CPow}
\end{equation}
Changes in the local power spectrum, such as those discussed above in the context of ring-diagram analysis, will affect the cross-covariance.

\subsection{Time-Distance Helioseismology}
Time-distance helioseismology \citep{Duvall1993,Kosovichev1996, Duvall1997} consists of (i)~measuring wave travel times from the cross-covariance function, and (ii)~inverting the travel times to infer the solar subsurface structure and flows.

As discussed in the {\bf Side Bar}, the cross-covariance $C(\br_1,\br_2,t)$ is closely related to a Green's function that gives the wave response at $(\br_2,t)$ to a source located at $(\br_1,t=0)$. Thus the cross-covariance is sensitive to the wave propagation conditions (structure and flows) between the two surface points $\br_1$ and $\br_2$. The sensitivity of the cross-covariance to any particular local change in the solar interior is a non-trivial research topic, because the local wavelength of solar oscillations is not necessarily small compared to the length scales of the heterogenities in the Sun. In addition, this sensitivity depends strongly on the combination of waves that contribute to the cross-covariance function. This will be discussed in Section~\ref{sec.forward_inverse}.

For the sake of simplicity, consider a constant horizontal flow $\bu$. According to Equations~\ref{eq.CPow} and~\ref{eq.pring}, the effect of such a flow is a Galilean translation of the unperturbed (no flow) cross-covariance, $C_0$, according to:
\begin{equation}
C(\br_1,\br_2,t) = C_0(\br_1, \br_2-\bu t, t) .
\end{equation}
In practise, this result is only an approximation because the power spectrum is often subject to additional filtering, which is not included in Equation~\ref{eq.CPow}
\citep{Gizon2002}.
The waves travel faster along the flow than against the flow. If $\bu$ is directed from $\br_1$ to $\br_2$, then the $t>0$ ridges of the time-distance diagram will be shifted to smaller time-lags $t$ and the $t<0$ ridges will be shifted to more negative $t$. Thus a flow will break the symmetry between the $t>0$ and $t<0$ parts of the cross-covariance. In contrast, a horizontally uniform change in, e.g., sound speed would introduce a time-symmetric change in the cross-covariance.

Several techniques have been proposed to measure travel times from the cross-covariance.
The (phase) travel times for inward- and outward-going waves are measured by fitting a Gabor wavelet to the two branches of the cross-covariance \citep{Duvall1997,Kosovichev1997SCORE}.
The travel times can also be measured with a simple one-parameter fit \citep{Gizon2002,Gizon2004APJ}, as is done in geophysics \citep[e.g.][]{Marquering1999}.

\begin{figure}[t!]
\centering
\includegraphics[width=\linewidth]{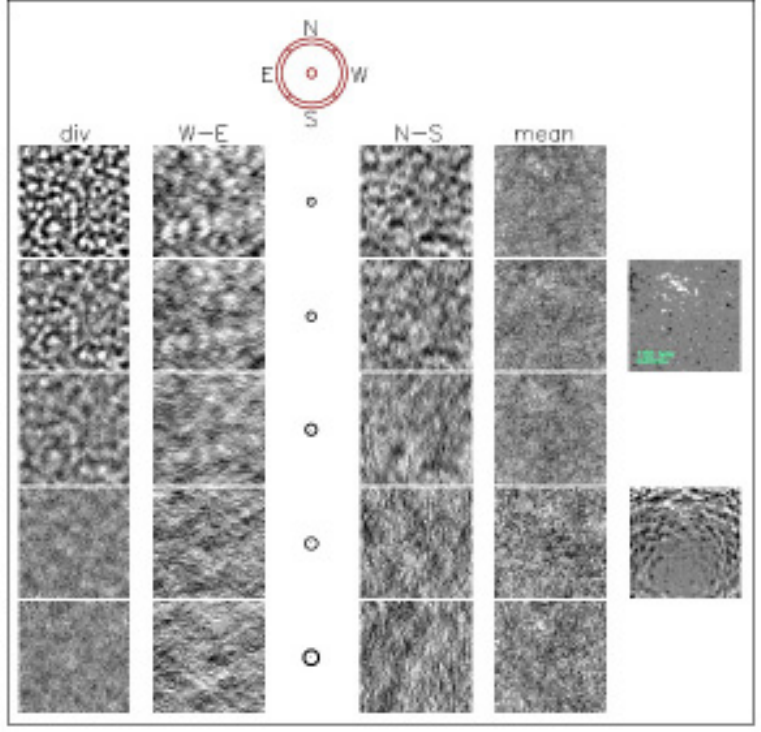}
\caption{
Maps of the travel times using the annulus/quadrant geometry and MDI/SOHO high-resolution data  \citep{Duvall1997}. The observation duration is $T=8.5$~hr. Each frame is 370~Mm on a side. Each row corresponds to a different annulus radius from 12~Mm to 27~Mm.  The columns show the four types of travel-time measurements. From left to right: (i) travel-time difference  between the outward-going waves and the inward-going waves, (ii) travel-time difference between the westward- and eastward-going  waves, (iii) travel-time difference between the northward- and southward-going waves, and (iv) mean of the travel times of the inward- and outward going waves.  The average magnetogram and the average Dopplergram are shown in the right most column. The travel-time perturbations are mostly due to the supergranular scale horizontal flows.
}
\label{fig.tt}
\end{figure}

The cross-covariance function computed between two spatial points is in
general very noisy. Spatial averaging is a useful tool to reducing random noise.
\citet{Duvall1997} considered an averaging scheme whereby the cross-covariance is computed between a point and a concentric annulus or quadrants of arc. For example, the cross-covariance between a point and an annulus
is used to study waves that propagate outward from the central point to the
annulus (positive time-lag) and inward (negative time-lag).
The difference between inward and outward travel times is sensitive to the horizontal divergence
of the local flow or to a local vertical flow, while the average travel time
is sensitive to the local wave speed.
Similarly, the covariance between a point and a north quadrant is used to
study waves that propagate either northward or southward. The various
combinations of travel times are shown in {\bf Figure~\ref{fig.tt}} for annulus radii ranging from $12$~Mm to $27$~Mm.
Each travel-time map is obtained by translating the  central point of the
annulus. With only a few hours of averaging, it is clear that
the travel-times differences are sensitive to the supergranulation flows.

Finally, the travel-time maps must be inverted (the inverse problem). This requires a model for the relationship between the travel-time perturbations and perturbations in solar properties (the forward problem). Recent progress regarding the interpretation of traveltimes is described in Section~\ref{sec.forward_inverse}.

\subsection{Helioseismic Holography}
Helioseismic holography \citep{Lindsey1997} and acoustic imaging
\citep{Chang1997}, which are virtually indistinguishable, are closely
related to the time-distance method.  In both of these methods,
observations of the wavefield (e.g.\  Doppler velocity) at the solar
surface are used to estimate the wavefield in the solar interior.
Separate estimates are constructed by computationally evolving the
observed wavefield either
forwards or backwards in time.  In helioseismic holography, these two
estimates are called the ingression ($H_-$, propagates forwards in time)
and egression ($H_+$,  propagates backwards in time) and are computed as
\begin{equation}
H_\pm(\br,z,\omega) = \int_{\mathcal P}\id^2\br' \; G_\pm
( \|\br'-\br \|, z, \omega) \phi(\br',\omega) ,
\end{equation}
where $\br$ is the horizontal focus position, $z$ is the focus depth,
and the integration over surface positions $\br'$ is carried out over
the region described by the "pupil" $\mathcal P$.  The functions $G_
\pm$ are  causal (subscript $-$) and anti-causal (subscript $+$)
Green's functions.  These Green's functions can be thought of as
propagators which (approximately) evolve the wavefield either
forwards or backwards in time.

The amplitude and phase of the correlation between the two estimates
$H_\pm$ contain information regarding wave propagation in the Sun \citep[e.g.][]
{Lindsey1997, Lindsey2000SCI}.
For example, in farside imaging (Section~\ref{sec.farside}), the phase of
the ingression-egression correlation is used to detect active regions.

\subsection{Direct Modeling}
Direct modeling \citep{Woodard2002, Woodard2007} is a method for interpreting correlations in the
wavefield. These
correlations, however, are not measured in real space but in the Fourier
domain. For example, any steady heterogeneity  in the Sun is expected to
introduce correlations between incoming and scattered waves with different
wave vectors but the same frequency.
Unlike time-distance helioseismology,
direct modeling can also treat time-varying perturbations, which couple waves
with different frequencies.  One of the main characteristics of direct
modeling, as its name suggests, is that it does not produce any intermediate data
products (e.g. no travel time map) as the inversions are carried out directly on the correlations.

\subsection{Fourier-Hankel Analysis}
\label{sec.hankel}
Fourier-Hankel analysis \citep{Braun1987} was specifically designed to study the wavefield
around sunspots. The analysis is carried out in a cylindrical coordinate system with origin centered on the
sunspot. The wavefield observed in an annular region around the sunspot is
decomposed into inward and outward propagating components, using a
Fourier-Hankel transform. The amplitudes and the phases of the incoming and
outgoing waves can be compared in order to characterize the interaction of the waves with the sunspot.
In particular Fourier-Hankel analysis was the first method to
measure the absorption coefficient of incoming waves by a sunspot
\citep{Braun1987}, defined by $(P_{\rm in}-P_{\rm out})/P_{\rm in}$, where
$P_{\rm in}$ and $P_{\rm out}$ are respectively the incoming and outgoing powers.
In addition, the Fourier-Hankel method has been used to  measure the phase
shift between the incoming and outgoing waves, as
well as the scattering from one mode to another \citep{Braun1995}.

\section{THE FORWARD AND INVERSE PROBLEMS}
\label{sec.forward_inverse}

\subsection{Weak Perturbation Approximation}
Many important solar features can be reasonably approximated, in the context of wave propagation, as small deviations from a horizontally uniform reference model. Examples include the supergranulation, meridional flow, torsional oscillations, and subsurface (but not surface) magnetic fields.  For this class of solar features, linear forward modeling can be employed.  The main advantage of linear forward models is that they lead to linear inverse problems, which is the only class of inverse problems that can be solved easily.

\subsubsection{\bf Sensitivity functions.}
In linear forward models, the helioseismic measurements (e.g., travel-time shifts) can be related to weak, steady, perturbations to a reference model through sensitivity functions (also called kernels) by equations of the form
\begin{equation}
d_i = \sum_\alpha\int_\odot K^\alpha_i(\bx) \delta q_\alpha(\bx) \id^3\bx \; ,
\label{eq.kernels}
\end{equation}
where the $d_i$ stand for an arbitrary set of helioseismic
measurements (for example travel-time perturbations $\delta\tau_i$), the
functions $\delta q_\alpha$ describe the deviations from the reference solar
model, and the functions $K^\alpha_i$ are the corresponding kernel functions.
The sum over $\alpha$ is over all types of perturbations to the solar model. A complete set of $q_\alpha$ includes, e.g.,
pressure, density, sound speed ($c$), flow velocity ($\bu$), and magnetic
field vector ($\bB$).
The integration variable $\bx$ is a three-dimensional position vector and runs
over the entire reference solar model.

\subsubsection{\bf The ray approximation.}
The ray approximation, in which the wavelength is approximated as small
compared to the other length scales in the problem (e.g., the scale heights of
the reference model, and the length scales of the perturbations $\delta q^\alpha$) has
been used extensively in time-distance helioseismology
\citep[e.g.][]{Kosovichev1996,Kosovichev1997SCORE}
to compute the travel-time sensitivity functions $K$.
A ray sensitivity kernel for the travel-time perturbation $\delta\tau(\br_1, \br_2)$, is zero everywhere except along the ray, $\Gamma$, that goes from $\br_1$ to $\br_2$  (see~Section\ref{sec.xc}).

The starting point is the local dispersion relation:
\begin{equation}
(\omega-\bk_{\rm t}\cdot\bu)^2 = c^2  k_{\rm t}^2 +\omega_{\rm ac}^2  ,
\end{equation}
where $\omega$ is the angular frequency, $\bk_{\rm t}$ is the three-dimensional total wave vector, $\bu$ is the vector flow, $c$ is the sound speed, and $\omega_{\rm ac}$ is the acoustic cut-off frequency.
In the ray approximation the travel time is given by the path integral of the phase slowness vector
$\bk_{\rm t}/\omega$. Under the assumption that $\omega$ is constant, the travel time perturbation is given by
\begin{equation}
\delta \tau = \frac{1}{\omega} \int_\Gamma \delta\bk_{\rm t} \cdot \hat{\bf n} \; \id s ,
%\qquad {\rm where}  \quad \delta \bk_{\rm t}\cdot \hat{\bf  n}  = - \frac{\omega}{c^2} \bu\cdot\hat{\bf n}   -  \frac{k_{\rm t}}{c} \delta c - \frac{\omega_{\rm ac}}{c^2 k_{\rm t}} \delta\omega_{\rm ac}
\end{equation}
where $\hat{\bf n}$ is the unit vector along $\Gamma$ and $\delta \bk_{\rm t}$ is the perturbation to the total wave vector caused by the flow $\bu$, the change in the sound speed, $\delta c$, and the change in the acoustic cut-off frequency, $\delta \omega_{\rm ac}$. Notice that the ray path, $\Gamma$, is assumed to be unchanged to first order. Although the ray approximation has played an important role in local helioseismology, it does not account for finite wavelength effects and other complications \citep[cf.][]{Bogdan1997,birch2009}.
\subsubsection{\bf The first Born approximation.}
In the first-order Born approximation, the perturbations $\delta q_\alpha$
cause a perturbation to the wave field that is due to single scattering only.
In this approximation, first-order finite wavelength effects, such as
Fresnel zones and wavefront healing, are included \citep[e.g.,][]{Hung2001}.
The Born approximation has
been used in the seismology of the Earth \citep[e.g., in the search for mantle
plumes by][]{Montelli2004}.
\citet{Gizon2002} give a detailed theoretical treatment of the application of
the Born approximation to problems in time-distance helioseismology.

Born kernels for the effects of sound-speed changes on travel times have been obtained by, e.g., \citet{Birch2000} and \citet{Jensen2000}. Born kernels for flows have been computed for time-distance helioseismology by \citet{Gizon2000}, \citet{Birch2007APJ}, and \citet{Jackiewicz2007APJ}. Three-dimensional sensitivity kernels for flows and  ring-diagram analysis are given by \citet{Birch2007AN}.
Although magnetic perturbations  are expected to be small just a few hundred km below the surface \citep{Gizon2006APJ}, corresponding kernels have not been obtained yet.

\begin{figure}[t!]
\centering
\includegraphics[width=13.cm]{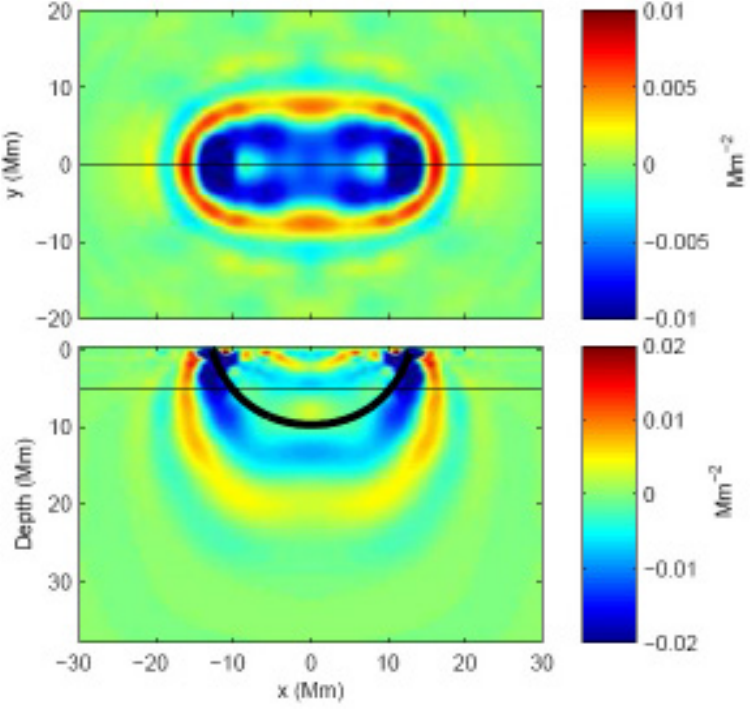}
\caption{
Linear sensitivity of a single-skip mean travel-time shift to local changes in the square of the sound speed. The observation points are located at $(x,y,z)=(\pm 12.5,0,-0.2)$~Mm. The top panel shows a horizontal slice at a depth of 5~Mm. The bottom panel shows a vertical slice at $y=0$. The heavy black line in the lower panel shows the single-skip ray path connecting the observation points. The kernel has been scaled with the background speed to enhance the visibility of the subsurface structure. The ringing away from the ray path is due to the band-limited nature of solar oscillations. The linear sensitivity is almost zero along the ray path, this a generic feature of sound-speed kernels in three dimensions, and is well known in seismology \citep{Marquering1999,Dahlen2000}.
}
\label{fig.kernel}
\end{figure}

{\bf Figure~\ref{fig.kernel}} shows slices through an example travel-time
sensitivity kernel for perturbations in the squared sound-speed
($\delta d_i = \delta\tau$ and $q_\alpha=c^2$ in Equation~\ref{eq.kernels}).  The main features are that the kernel is zero along the ray
path, is maximum (absolute value) in a shell around the ray path, and shows
substantial ringing.   These kernels have been called ``banana-doughnut
kernels'' in the context of seismology of the Earth \citep{Marquering1999}.%,Dahlen2000}.
The zero along the ray path is due to the lack of a geometrical delay (change in path length) for small scatterers located on the reference ray path. In the solar case, there are additional hyperbolic features across the ray path, due to the presence of distant sources \citep{Gizon2002}.

The first-order Born approximation have been tested using exact solutions \citep[e.g.][]{Gizon2006APJ} and numerical simulations \citep[e.g.][]{Hung2000,  Birch2001, Baig2003, Birch2004Felder} for simple cases.
\citet{Duvall2006} used solar observations to construct two-dimensional travel-time kernels using small (sub-wavelength) magnetic features as point scatterers.

\subsubsection{\bf Inversions and Resolution kernels.}
Linear inversions have been developed for ring-diagram analysis
and time-distance helioseismology, based on experience
gained from global helioseismology \citep[e.g.,][and references therein]{JCD1990}.
Two different inversion procedures
are commonly used.

Let us consider 1D depth inversions for the sake of simplicity.
The first inversion method, called Regularized Least Squares
\citep[RLS, e.g.,][]{Kosovichev1996,Haber2000}, is simply
a fit to the observational data $\delta d_i$ under conditions of smoothness.
The second inversion method, called Optimally Localized Averaging
\citep[OLA,][]{Haber2004}, looks for a linear combination
of the kernels (an averaging kernel)
that is spatially localized around a target depth, $-z_0$.
Both methods build averaging kernels of the form
\begin{equation}
{\cal K}(z;z_0) = \sum_i c_i(z_0) K^\alpha_i(z)
\end{equation}
where the $c_i$ are coefficients to be determined. A regularization is applied
to ensure that the error on the inferred $q_\alpha$ is not too large, or that $q_\alpha$
is smooth. In addition, the cross-talk between the inferred
 $q_\alpha$ and all other quantities $q_{\beta}$, $\beta\neq\alpha$,
 should be avoided.
Inversions require a good knowledge of
the noise covariance matrix of the measurements, which
can be estimated directly form the data (spatial averaging) or from a model
\citep{Gizon2004APJ}.

\begin{figure}[t!]
\includegraphics[width=\linewidth]{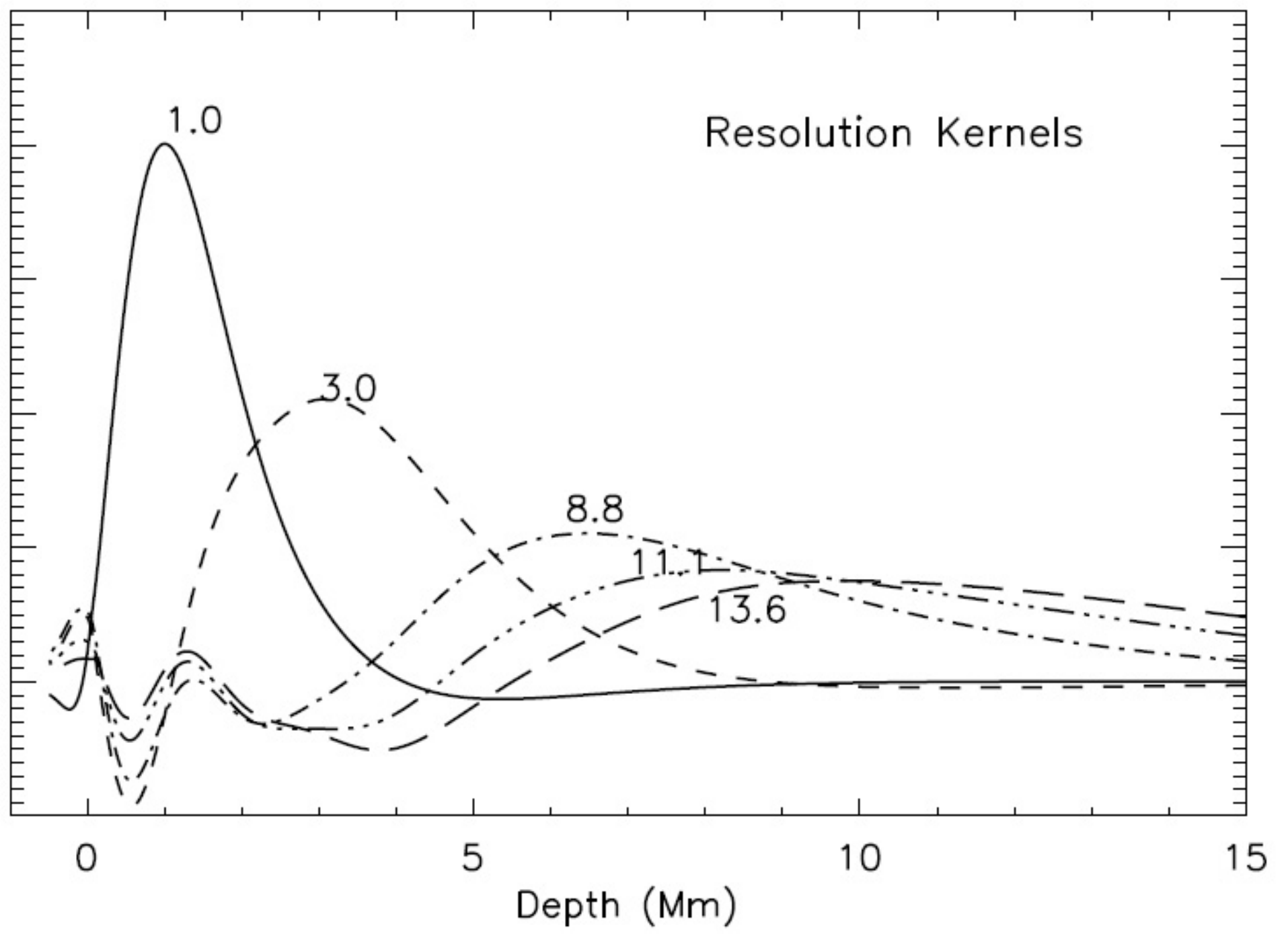}
\caption{
Averaging kernels, ${\cal K}$, from 1D RLS inversions for ring-diagram analysis (depth sensitivity to horizontal flows). The numbers (in units of Mm) refer to the average depths of the averaging kernels. Courtesy of Irene Gonz\'alez-Hern\'andez.
%{\bf NOTE TO EDITOR: remove `Resolution Kernels' from figure and replace y-axis label by `Averaging kernels'.}
}
\label{fig.ola_kernels}
\end{figure}

The RLS and OLA methods give similar answers, although
the RLS averaging kernels are perhaps more likely to have undesired sidelobes
near the surface \citep{Haber2004}.  {\bf Figure~\ref{fig.ola_kernels}} shows example RLS
averaging kernels for ring-diagram analysis, in the near-surface layers.
These particular kernels are used to infer horizontal flows.
Very similar averaging kernels are obtained for time-distance helioseismology
 \citep[e.g.][]{Jackiewicz2008SOL}.

Most time-distance inversions assume that the kernels are invariant by horizontal
translation, so that the horizontal convolution of the kernels with the
$\delta q_\alpha$ becomes a multiplication in Fourier space. This property is
used to speed-up the inversions \citep{Jacobsen1999}.

Recent progress includes the inclusion of the Born kernels in the
inversions and the noise correlations \citep[e.g.][]{Couvidat2005}.
An OLA inversion for the horizontal and vertical components of the flows
was implemented by \citet{Jackiewicz2008SOL}, in which
the Born kernels and the noise covariance matrix are both consistent with the
definition of the observed travel times.
We note that the vertical component of velocity has been indirectly estimated from ring-diagram inversion by requiring mass conservation \citep[e.g.][]{Komm2004}.

\subsection{Strong Perturbation Regime}
\label{sec.strong_perturbation}
The solar atmosphere is permeated by concentrations of magnetic field with strength $B > 1$~kG. This magnetic field profoundly affects the solar atmosphere as well as the solar oscillations in the upper layers \citep[][and references therein]{Bogdan1995}. The effects on the waves are not small and it is formally not justified to employ a single-scattering approximation to describe them.  For example, the first Born approximation is not expected to capture the interaction of f and p modes with the near-surface layers of sunspots \citep[e.g.][]{Gizon2006APJ}.

The Lorentz force is an additional restoring force that permits the existence of new oscillation modes. In the case of a spatially uniform model with no gravity, it is possible to identify three types of magneto-hydrodynamics (MHD) waves: the fast, slow, and Alfv\'en waves. In gravitationally stratified magnetized atmospheres, this classification can only be applied locally. The waves couple near the layer where the sound speed and the Alfv\'en speed, $a=B/\sqrt{4\pi\rho}$, are equal \citep[e.g.][]{Schunker2006}. In a typical sunspot, the $a=c$ surface is only a few hundred km below the quiet-Sun photosphere.

\begin{figure}[t!]
\begin{center}
\includegraphics[width=11.5cm]{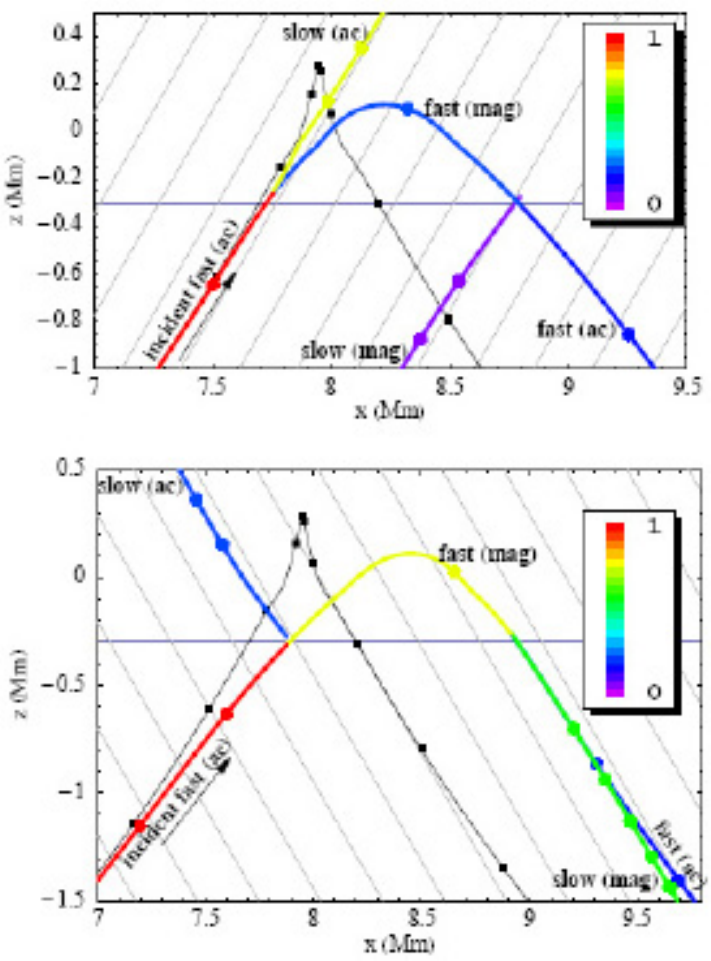}
\end{center}
\caption{Ray paths in model surface layers with 2~kG uniform magnetic field inclined at $\pm 30^\circ$ respectively to the vertical, as shown by the background gray lines. The incoming 5~mHz rays (arrows) have lower turning points at $z=-5$~Mm and are shown in red. The two frames correspond to two different attack angles. The horizontal gray line indicates where the sound speed and the Alfv\'en speed coincide, which is approximately where mode conversion happens. The fractional energy remaining in each resulting ray is indicated by the color legend. The dots on the ray paths indicate 1~min group travel time intervals. The thin black curve represents the acoustic ray that would be there in the absence of magnetic field. Note that the fast ray is faster through the surface layers than the non-magnetic ray. Figure and caption from \citet{Cally2007}.
Figure copyright {\it Wiley-VCH Verlag GmbH \& Co.\ KGaA}. Reproduced with permission.}
\label{fig.cally}
\end{figure}

The ray approximation can be extended in order to study wave propagation in MHD problems \citep[e.g.][]{Schunker2006}. The magnetic field affects travel times, mode frequencies, and amplitudes. These effects depend sensitively on the geometry, in particular the angle between the incident wave vector and the magnetic field vector at the $a=c$ layer.  {\bf Figure~\ref{fig.cally}} shows two example ray calculations for the case of an incoming acoustic wave approaching the solar surface from below (the magnetic field, wave vector, and gravity are in the same plane).  In these two calculations all of the parameters are the same except that the angle of the magnetic field has been changed. In {\bf Figure~\ref{fig.cally}}{\boldmath $a$}, the wavevector of the acoustic wave is nearly aligned with the magnetic field at the $a=c$ layer.  In this case,  most of the energy is transmitted to the acoustic (slow) mode of the $a>c$ region.   This acoustic mode escapes along the magnetic field lines (ramp effect).   Some energy is, however, converted to the fast (magnetic) mode which is then refracted by the increase of the Alfv\'en speed with height.
In {\bf Figure~\ref{fig.cally}}{\boldmath $b$}, the incident wavevector makes a large angle with the magnetic field. At the $a=c$ level, the acoustic wave converts mostly into a magnetic (fast) mode that is refracted back into the Sun and then mostly becomes a downward propagating magnetic (slow) mode, while a small fraction of energy continues as a fast mode. The slow mode is not seen again at the surface, and thus removes energy from the surface wavefield. {\bf Supplemental Movies 3\,--\,6} illustrate generalized ray theory for various values of the attack angle between the wave vector and the magnetic field.
We note that in three dimensions, where the magnetic field, wave vector, and gravity are not all co-planar, strong coupling to the Aflv\'en wave also occurs \citep{CallyGoossens2008}.

Numerical simulations are an important tool to study waves in magnetized regions and sunspots. Two different approaches are employed. The first approach is numerical simulations of wave propagation through prescribed background models \citep[e.g.][]{Cally1997,Khomenko2006,Cameron2007,Hanasoge2008,Parchevsky2009}. This approach permits the study of wave propagation without the complications of solving for convection and it gives the freedom to choose various background models. Typically, these codes solve the equations of motion for small-amplitude waves. The second approach is realistic numerical simulations of magnetoconvection \citep[e.g.][]{Rempel2009}. Such simulations include self-excited waves and aim at simulating realistic solar active regions. This approach is very promising, but computer intensive.

\begin{figure}[t!]
\includegraphics[width=10.cm]{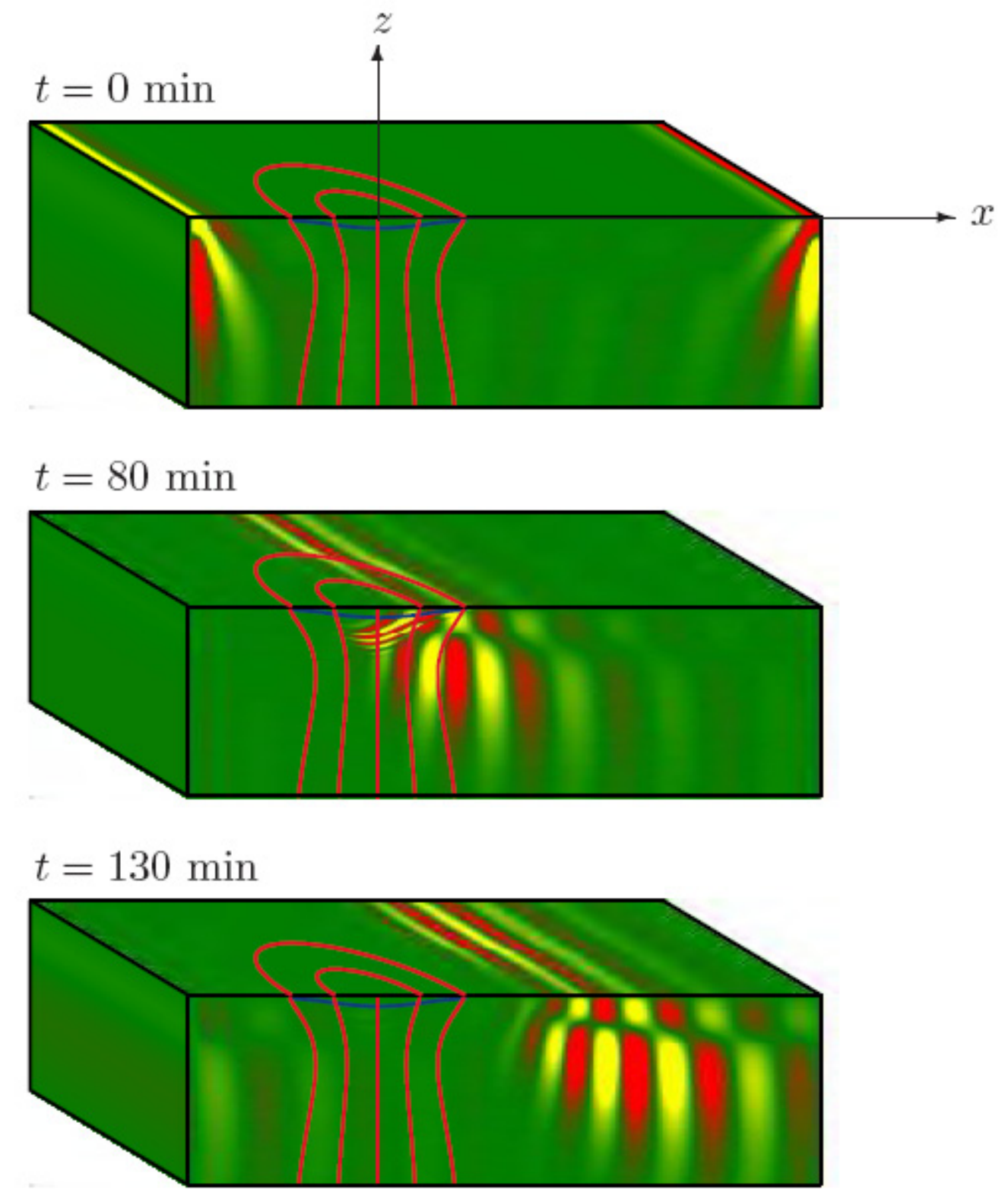}
\caption{Numerical simulation of the propagation of a p$_1$ plane wave packet through a model sunspot using the MHD code of \citet{Cameron2007}. The  sunspot model is similar to the model of \citet{Schlueter1958} with a maximum vertical field $B_z=3$~kG at the surface; it is embedded in a Model S background atmosphere, stabilized with respect to convection. The wave packet is initially located 40~Mm to the left of the sunspot and propagates to the right.  The $x$-component of velocity is shown at times $t=0$, $80$, and $130$~min (positive values in red, negative values in yellow). The $a=c$ level is represented by the blue curve. The computational domain is $-40~\mbox{Mm}<x<105~\mbox{Mm}$, $-36.5~\mbox{Mm}<y<36.5~\mbox{Mm}$ and $-12.5~\mbox{Mm}<z<1.5~\mbox{Mm}$. Only half of the box ($y>0$) is shown. The boundary conditions are periodic in the horizontal directions and there are two sponge layers (not shown here) at the top and at the bottom of the box to avoid the reflection of the waves back into the computational domain. The $t=80$~min snapshot clearly shows the slow magneto-acoustic waves propagating down the sunspot.  Because these slow waves are transverse, they are easily seen as oscillations in the $x$-component of velocity. See {\bf Supplemental Movie 8}.}
\label{fig.slim_vx}
\end{figure}

{\bf Figure~\ref{fig.slim_vx}} shows an example calculation of the propagation of a p$_1$ wave packet through a simple sunspot model using the three-dimensional code of \citet{Cameron2007}. This example shows that the transmitted wave packet is phase-shifted by the sunspot (increased effective wave speed) and has a reduced amplitude compared to the quiet-Sun value. Also seen is the partial conversion of incoming p modes into downgoing slow modes. This process is responsible for absorption of acoustic energy by sunspots \citep[][and Section~\ref{sec.sunspots}]{Spruit1992,Cally1997,Crouch2005SOL}. An f mode wave packet is affected in a similar fashion \citep{Cally1997,Cameron2008}.

The problem of inferring the subsurface structure and dynamics of solar active regions is a difficult one. In principle, standard linear inversions cannot be used because surface magnetic perturbations are not small with respect to a quiet Sun reference model.
No non-linear inversion has been implemented yet. Instead there have been attempts to construct simple parametric models of magnetic regions, which have a helioseismic signature that is consistent with the observations. This approach does not require that perturbations be small. Forward models of sunspots have been proposed by, e.g., \citet{Crouch2005} and \citet{Cameron2008}; they will be discussed in more detail in Section~\ref{sec.sunspot_structure}.

Now that we have some understanding of the methods of local helioseismology and their diagnostic potential, we turn to a description of the main observational results: near-surface convection (Section~\ref{sec.superg}), sunspots (Section~\ref{sec.sunspots}), extended flows around active regions (Section~\ref{sec.SSW}), global scales (Section~\ref{sec.global_scales}), farside imaging (Section~\ref{sec.farside}), and flare-excited waves (Section~\ref{sec.flares}).

\section{NEAR-SURFACE CONVECTION}
\label{sec.superg}

\subsection{Supergranulation and Magnetic Network}

Solar supergranules are horizontal outflows with a typical size of 30~Mm,
outlined by the chromospheric network \citep[e.g.][]{Leighton1962}.
They have horizontal velocities of order $200$~\ms
and lifetimes of one to two days.

 \citet{Duvall1997} found that p-mode  travel times contain information at supergranular length scales ({\bf Figure~\ref{fig.tt}}). As a demonstration of this, the line-of-sight component of velocity, estimated from the travel times assuming that vertical motions are negligible, was found to be highly correlated with the time averaged Dopplergram, thus confirming that local helioseismology is capable of probing convective flows at supergranulation length scales.
\citet{Duvall2000} extended the analysis to f modes to infer horizontal flows within 2 Mm below the surface. Because they propagate horizontally, f modes are well suited to measure horizontal flows and their horizontal divergence. The flows from f-mode time-distance helioseismology compare well with flows estimated from local correlation tracking of mesogranulation \citep{DeRosa2000}. Recently, \citet{Woodard2009} has shown that direct modeling can be used to detect convection on scales of space and time that are smaller than those of supergranulation.

\begin{figure}[t!]
\includegraphics[width=\linewidth]{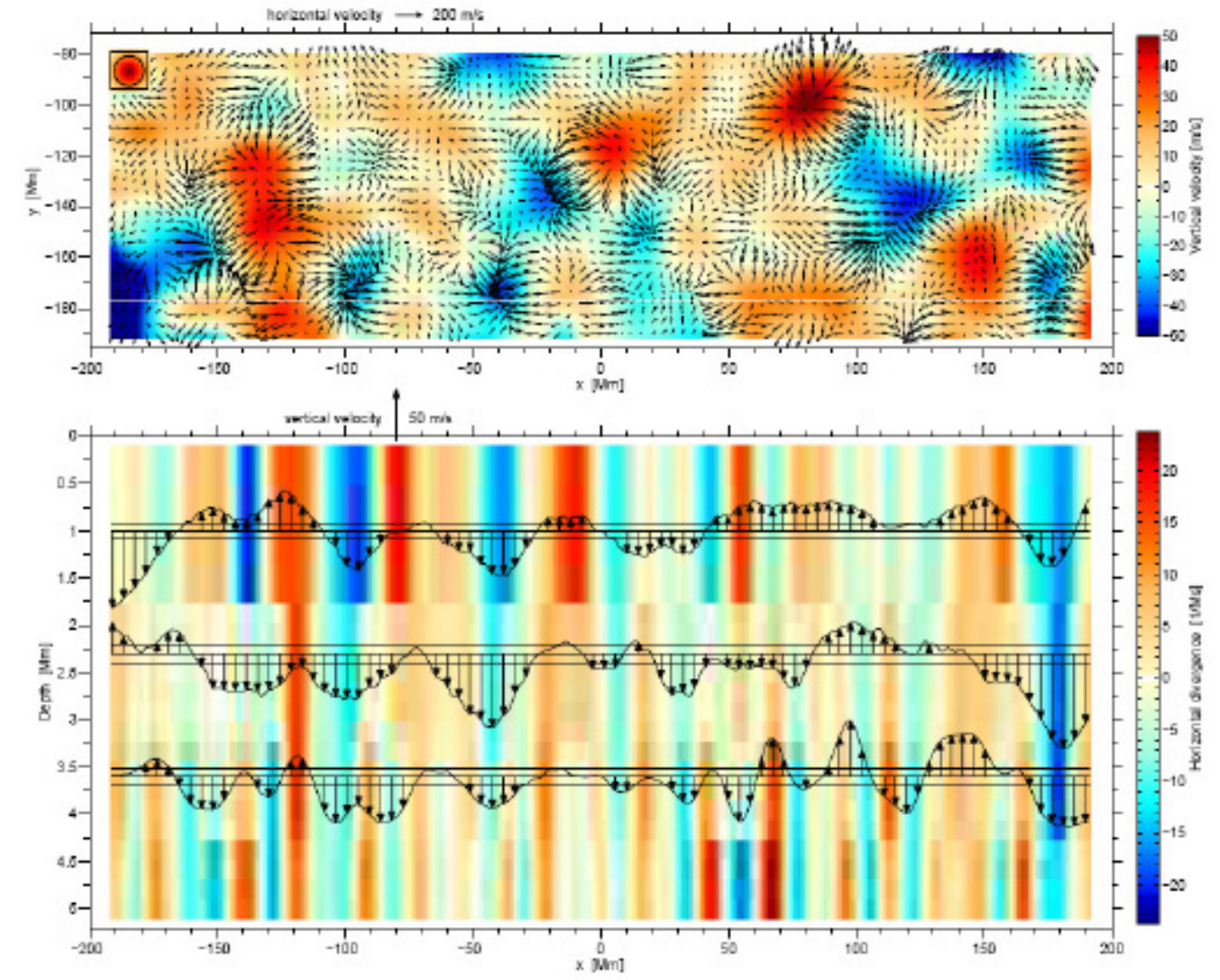}
\caption{
Inversion for vector flows in the near surface layers. The travel times were measured for ridges f and p$_1$ through p$_4$ and inverted using an OLA technique. The observation time is $T=3$~days. {\it Top panel}: Horizontal slice at the depth of $1$~Mm showing the two horizontal components ($u_x$ and  $u_y$, arrows)  and the vertical component ($u_z$, colors) of the vector flow field. The FWHM of the averaging kernel is 10~Mm (inset in top-left corner).  The most visible features correspond to long-lived supergranules. {\it Bottom panel}: Vertical slice at $y=-176$~Mm (white line in $a$) showing the horizontal divergence of the flow field as a function of $x$ and depth (colors) from the same 2+1D inversion. The vertical arrows show the vertical velocity $u_z$ for the three  ridges f, p$_1$, and p$_2$ (from top to bottom) from a 2D horizontal inversion, plotted at the depths corresponding to the mean depth of the respective averaging kernels. The one-sigma level of random noise in $u_z$ is equal to 10~\ms by construction (horizontal black lines). Adapted from \citet{Jackiewicz2008SOL}.
}
\label{fig.vz_combo}
\end{figure}

{\bf Figure~\ref{fig.vz_combo}} shows the most recent inversion of travel-times \citep{Jackiewicz2008SOL} using modes f through p$_4$. This inversion incorporates a full treatment of finite-wavelength effects (first-order Born approximation), modeling of the details of the measurement procedure, and a treatment of the statistical properties of noise. The vector flow field, averaged over $T=3$~days, is dominated by long-lived supergranules. As seen in the figure, the divergent flows are co-spatial with upflows with about 30~\ms rms velocity (with maximum values of $\sim 50$~\ms). Near the surface, the vertical velocity can be measured in supergranules with a noise level of about 10~\ms for 24~hr averages and a horizontal resolution of about 10~Mm. Estimates of the vertical velocity in supergranules from direct Doppler measurements can only be obtained near disk center and are in the range $10$\,--\,$30$~\ms \citep[][and references therein]{Hathaway2002}, which is consistent with the results of local helioseismology.

Because noise reduction requires time averaging, the finite lifetime of supergranulation implies a strict limitation on the maximum depth at which we can probe the flow field before it evolves
substantially. Using the f and p$_1$--p$_4$ modes, \citet{Woodard2007} found that random noise dominates below about 4~Mm. Probing supergranules at greater depth involves statistical analysis over large fields of view and many supergranulation lifetimes in order to reduce the noise: this allows the study of the average properties of the flows at depth. Inversions of convective flows tens of Mm below the surface are challenging as they require excellent models of the sensitivity of travel times to subsurface flows \citep[see][for a discussion]{Braun2003} and claims of the detection of a supergranulation return flow are apparently inconsistent \cite[e.g.][]{Duvall1998,Zhao2003}.

\begin{figure}[t!]
\includegraphics[width=\linewidth]{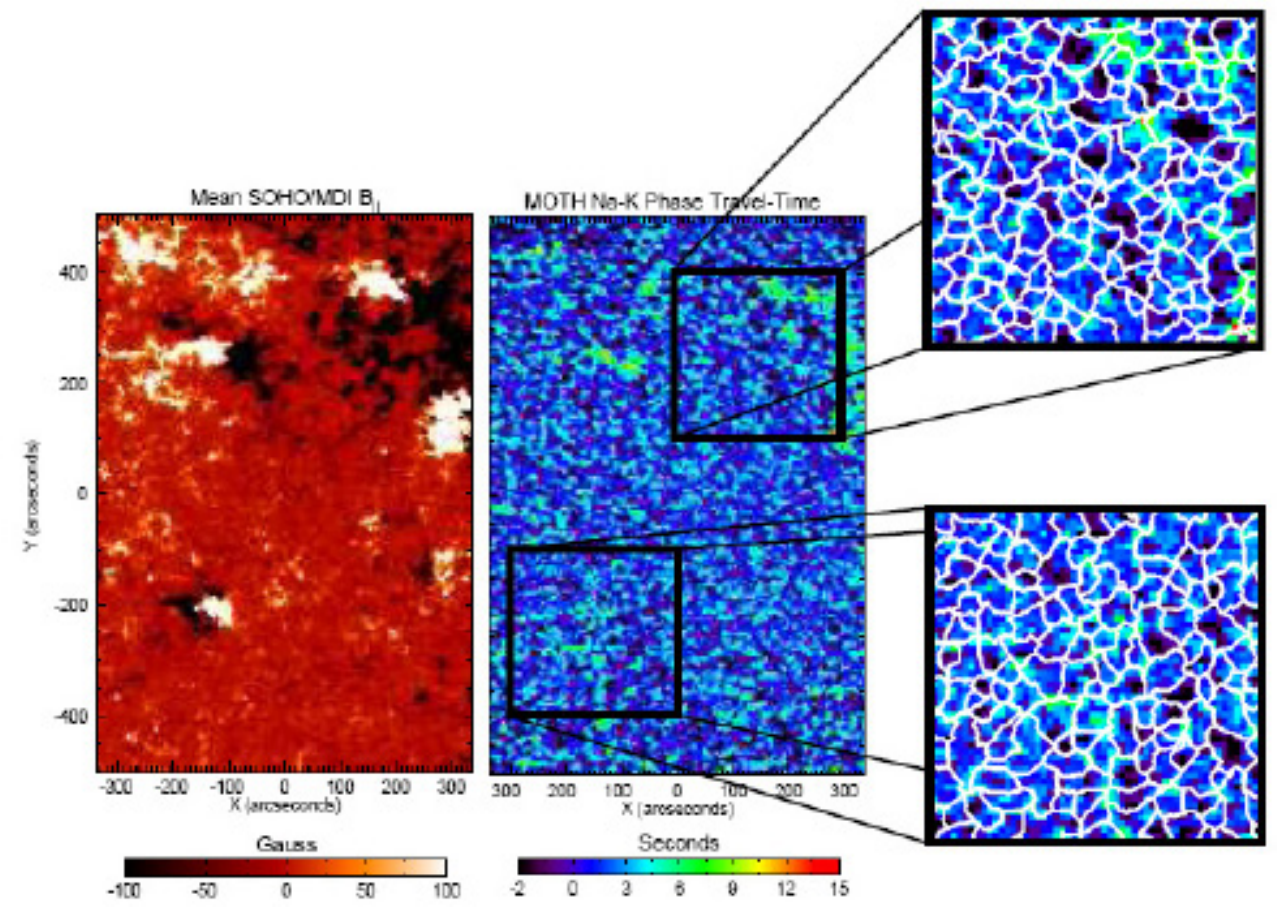}
\caption{
{\it (Left)} Map of the average MDI/SOHO line-of-sight component of the magnetic field in the Sun's photosphere for the 107 hr period starting 06:59 UT on 2003 January 6. {\it (Middle)} Map of phase travel time \citep{Finsterle2004APJL} for magnetoacoustic waves with frequencies near 3 mHz based on contemporaneous, simultaneous Doppler velocity data of the full solar disk as viewed at 5890~\AA\ (Na) and 7699 \AA\ (K). {\it (Right)} Magnified views of two regions of the phase-travel-time map overlaid with an estimate of the location of the boundaries of the supergranular-scale convective cells as determined using a segmentation of the mean intensity image at 5890~\AA.
Note that there is not a significant travel-time signal in all of the observed plages, only in regions where the field is highly inclined. This signal is noticeably larger than that in the boundaries of the supergranules. This is probably due to the larger magnetic filling factor in the plage. Figure and caption from \citet{Jefferies2006}.
}
\label{fig.portals}
\end{figure}

The pattern of divergent flows in the surface layers is outlined by a network of small
magnetic features \citep[see][and {\bf Supplemental Movie 7}]{Duvall2000,Braun2003}. This can be understood as the magnetic field is swept by the flows and concentrates at the boundaries of supergranules \citep[e.g.][]{Galloway1977}.
The connections between the magnetic network and the propagation behavior of acoustic waves
in the solar chromosphere can be studied by cross-correlating the observations  of solar oscillations
at multiple heights in the solar atmosphere \citep{Finsterle2004APJL}.
\citet{Jefferies2006} showed that inclined magnetic field lines at the boundaries of supergranules provide `portals' through which low-frequency ($<5$~mHz) slow MAG waves can propagate up into the solar chromosphere (see {\bf Figure~\ref{fig.portals}}). This is because the cut-off frequency is lowered in the magnetic network relative to the quiet Sun by a factor $\cos\theta$, where $\theta$ is the inclination of the magnetic field to the vertical. These low-frequency upward traveling waves have been suggested to act as a source of heating of the quiet-Sun chromosphere -- as an alternative to Joule heating due to magnetic field reconnection or mechanical heating due to high-frequency waves.

\subsection{Rotation-Induced Vorticity}
\begin{figure}[t!]
\includegraphics[width=\linewidth]{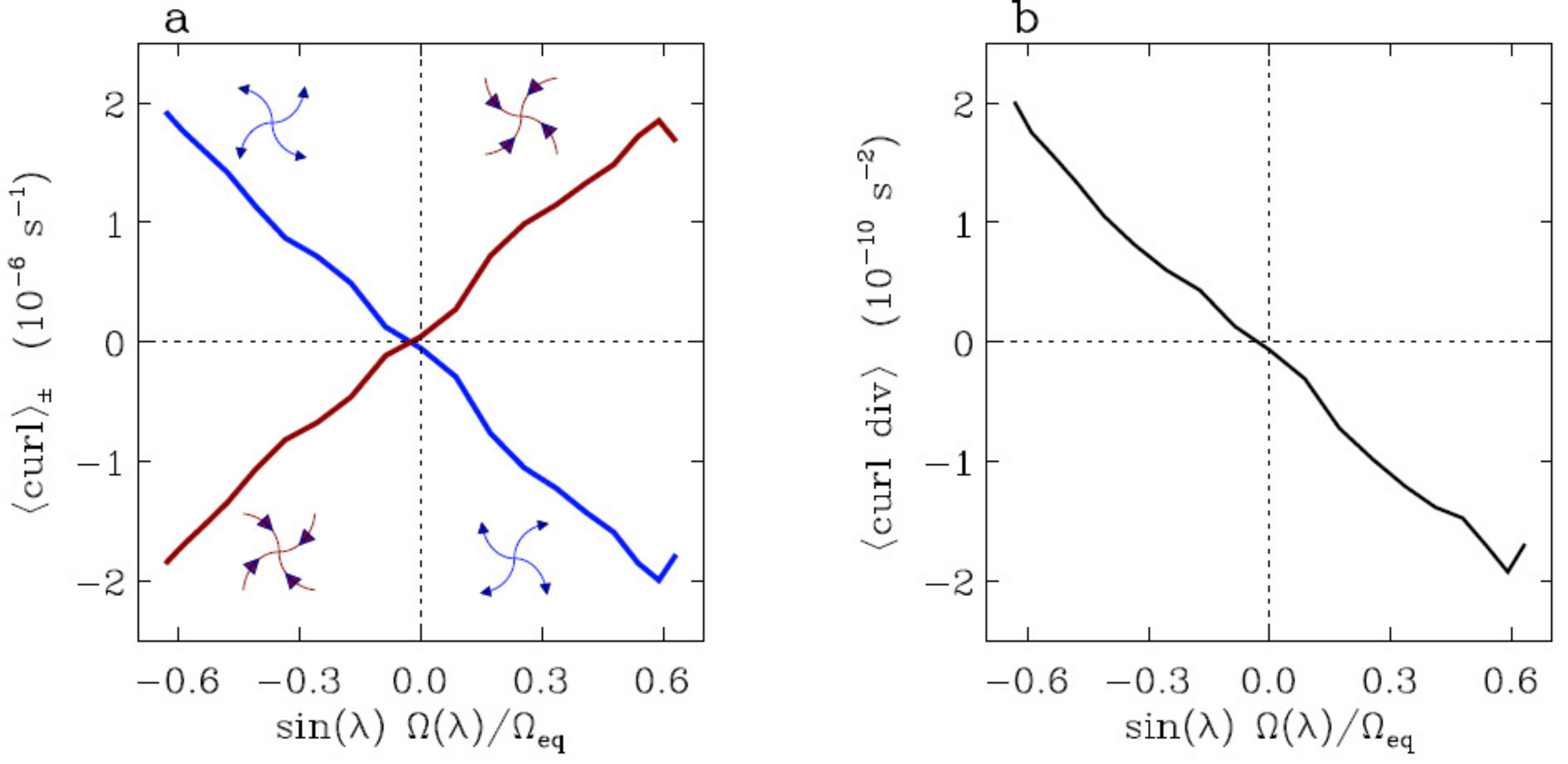}
\caption{Effect of the Coriolis force on supergranulation flows.
({\it a}) Vertical vorticity (curl) averaged over regions with positive horizontal divergence ($\langle {\rm curl} \rangle_+$, blue curve) and negative horizontal divergence ($\langle {\rm curl} \rangle_-$, red curve) as functions of $\sin(\lambda) \Omega(\lambda)/\Omega_{\rm eq}$, where $\lambda$ is the heliographic latitude and $\Omega/\Omega_{\rm eq}$ is the local surface angular velocity relative to the equator. A vorticity of 1~M\,s$^{-1}$ corresponds to an angular velocity of $2.5^\circ$~day$^{-1}$ or a typical tangential velocity of 10~\ms.
({\it b}) Horizontal average $\langle {\rm curl} \; {\rm div} \rangle$ versus $\sin(\lambda) \Omega(\lambda)/\Omega_{\rm eq}$.
Adapted from \citet{Gizon2003ESA}.
}
\label{fig.coriolis}
\end{figure}

Rotation is expected to have a small effect on the dynamics of the
supergranulation through the Coriolis force. As solar convection is highly
turbulent, this effect can only be studied in a statistical sense using
several months of data. For example, in the northern hemisphere, divergent
flows are expected to have a slight positive correlation with clockwise
vertical vorticity. \citet{Duvall2000} and \citet{Gizon2003ESA} used
time-distance helioseismology to make the first measurement of this small
effect. After removing the average rotation and meridional circulation from the  inferred flows, \citet{Gizon2003ESA} studied the relationship between the horizontal divergence of  the flows, denoted by `div', and the vertical component of vorticity, denoted by `curl'.  {\bf Figure~\ref{fig.coriolis}}{\boldmath $a$} shows the latitudinal dependence of $\langle {\rm curl}\rangle_+$ and $\langle {\rm  curl}\rangle_-$, respectively defined as the averages of the curl over regions of positive and negative div.  In the northern hemisphere, diverging flows preferentially rotate clockwise, whereas converging flows preferentially rotate counter-clockwise. This pattern is reversed in the southern hemisphere.  This situation is an expected consequence of the Coriolis force. Furthermore, the latitudinal dependence of $\langle {\rm  curl}\rangle_+$ and $\langle {\rm  curl}\rangle_-$ are observed to be nearly exactly proportional to the radial component of the solar angular velocity, $\sin(\lambda) \Omega(\lambda)$, where $\lambda$ is latitude and $\Omega$ is the solar angular velocity.

{\bf Figure~\ref{fig.coriolis}}{\boldmath $b$} shows that the average of the product of div by  curl is given by
\begin{equation}
\langle {\rm div} \, {\rm curl} \rangle \simeq -3\times 10^{-10}\; \sin(\lambda) \Omega(\lambda)/\Omega_{\rm eq} \; {\rm s}^{-2} \; ,
\end{equation}
where $\Omega_{\rm eq}$ is the equatorial angular velocity.
Simple dimensional analysis of the equations of motion
predicts that $\langle {\rm div} \, {\rm curl}
\rangle \sim -{\rm Co}(\lambda)/\tau^2$ where ${\rm
  Co}(\lambda)=2\tau\Omega(\lambda)\sin{(\lambda)}$ is the local Coriolis
number and $\tau$ is the characteristic correlation time of the turbulence.
For example, $\tau=2$~days implies ${\rm Co}(\lambda)/\tau^2 \sim 3\times
10^{-11} \sin(\lambda) \Omega(\lambda)/\Omega_{\rm eq} \; {\rm s}^{-2} $.
  It is not clear if this difference of a factor of ten  indicates missing physics or simply reflects the uncertainty in such estimates.  Attempts at improved quantitative results suffer from arbitrarily tunable parameters.

Cyclonic convection is a means to generate poloidal field from toroidal field
and is thus important in many dynamo models \citep[for a recent review
see][]{Charbonneau2005}.  In these models, the sign of the kinetic helicity
${\cal H}_{\rm kin} = \langle \bu \cdot \left(\nabla\wedge \bu \right)
\rangle$ determines the strength (and sign) of this effect.  Helioseismic
measurements imply that the kinetic helicity at supergranulation scales is negative in the northern hemisphere (and positive in the south).  This is an estimate rather than a direct measurement because the horizontal components of the vorticity have not yet been measured directly.

\subsection{Evolution of Supergranulation Pattern}
\citet{Gizon2003NAT} studied the Fourier spectrum of long time series of maps
of the horizontal divergence of the flows at supergranulation scales,
measured using f-mode time-distance helioseismology.
The observations reveal surprising characteristics: the signal has wavelike
properties (period around 6 days) and power is anisotropic (excess power in
the prograde and equatorward directions).
These observations have been confirmed independently by Zhao (private communication) and Braun
(private communication) using p-mode helioseismology.
The power peaks at a non-zero frequency that increases slightly with horizontal wavenumber.
Measurements of the Doppler shift of this apparent dispersion relation
has provided a robust method for measuring the rotation and meridional flow of
the solar plasma  \citep{Gizon2003NAT, Gizon2008}.
An interesting aspect of this work is that the inferred rotation and the meridional flow match the motion of the small magnetic features \citep[e.g.][]{Komm1993ROT, Komm1993MER}. On the other hand, correlation tracking measurements applied to the divergence maps overestimate rotation and underestimate the meridional flow by large amounts \citep[see][]{Gizon2005}. The time evolution of the supergranulation pattern does not reflect its advection by the plasma flow, although the two can be decoupled in Fourier space.

We note that \citet{Hathaway2006} demonstrated that the local correlation tracking of Doppler features on the Sun gives biased estimates of the rotation rate because of line-of-sight projection effects.
This case, however, is not directly comparable to the observations described above
since helioseismic divergence maps are not expected to be sensitive to line-of-sight projection effects at supergranulation scales.

The helioseismic observations of the wavelike properties of supergranulation
are still calling for an explanation.
%, either a physical explanation or some unidentified systematic error.
Supergranulation may perhaps be related to the traveling convection modes seen in idealized systems with rotation \citep[e.g.][]{Busse2007}.

\section{SUNSPOTS} \label{sec.sunspots}

In this section we discuss inferred flows in the immediate vicinity of sunspots (Section~\ref{sec.sunspot_flows}), the absorption of waves by sunspots (Section~\ref{sec.sunspot_sinks}), and the subsurface structure of sunpots (Section~\ref{sec.sunspot_structure}).  Recent reviews about sunspots are provided by, e.g.,  \citet{Solanki2003} and Moradi et al. (2009, submitted).

The absence of a sufficiently conclusive theory has allowed a wide range of ideas about the origin and structure of sunspots to develop. These range all the way from intuitive ideas directly inspired by the abundant observational clues, to mathematically oriented ones that require ignoring almost all of these clues. Some of the ideas should become testable if they make relevant predictions for the relatively shallow layers below the surface that are accessible to local helioseismology methods.

\subsection{The Anchoring Problem}
The magnetic forces exerted by the spot on its surroundings are significant. If it were not in a quasi-stable equilibrium in its observable layers, a spot would evolve on the time for the Alfv\'en speed to cross the size of the spot  (on the order of an hour), much shorter than the observed life times of spots (days to weeks).  The magnetic forces also make the sunspot plasma buoyant. Together, this gives rise to an 'anchoring problem' \citep[cf.][]{Parker1979BOOK}.
A sunspot cannot be just a surface phenomenon since magnetic field lines have no ends. The sunspot's field lines  continue below the surface. In contrast with a scalar field like pressure, the magnetic field of a sunspot cannot be kept in equilibrium simply by pressure balance at the surface: the tension in the magnetic field lines continuing below the surface exerts forces as well. The magnetic tension acting at the base of the spot keeps it together and prevents buoyancy from spreading it like an oil slick over the solar surface. Sunspots also rotate faster than the solar surface, indicating that they sense the increase of rotation with depth.

The question of sunspot equilibrium thus involves deeper layers, down to wherever the field lines continue. At which depth and by which agent is the sunspot flux bundle kept together? A very stable location is the boundary of the convection zone with the stably stratified radiative interior of the Sun. A layer of magnetic field floating on this boundary becomes unstable only at a field strength of about $10^5$G \citep{Schuessler1994}. The existence of such a critical field strength was hypothesized by \citet{Babcock1961}. The subsequent rise to the surface is what creates the observed bipolar active regions, as proposed by \citet{Cowling1953}. The action of the Coriolis force on flows in the magnetic field associated with the instability produces the poloidal field of the next cycle, and is observable on the surface in the form of the systematic tilt of active region axes with respect to the azimuthal direction \citep{Leighton1969}. A boost of confidence has been provided by recent realistic 3D radiative MHD simulations of the last stages of the emergence process of magnetic fields at the surface. These are beginning to look much like real observations \citep{Cheung2008}. Though largely qualitative, the view of the solar cycle developed by Babcock and Leighton appears to be the most fruitful frame of reference for interpreting the solar cycle.

\subsection{Moat Flow}\label{sec.sunspot_flows}

In the photosphere, sunspots are typically surrounded by diverging horizontal
outflows, termed moat flows, with amplitudes of several hundred \ms. These outflows typically extend to about twice the radius of the penumbra.   Moat flows were first detected using direct Doppler measurements \citep{Sheeley1972} and also can be inferred from the motion of small magnetic features \citep[e.g.][]{Brickhouse1988}.

\begin{figure}[t!]
\begin{center}
\includegraphics[width=\linewidth]{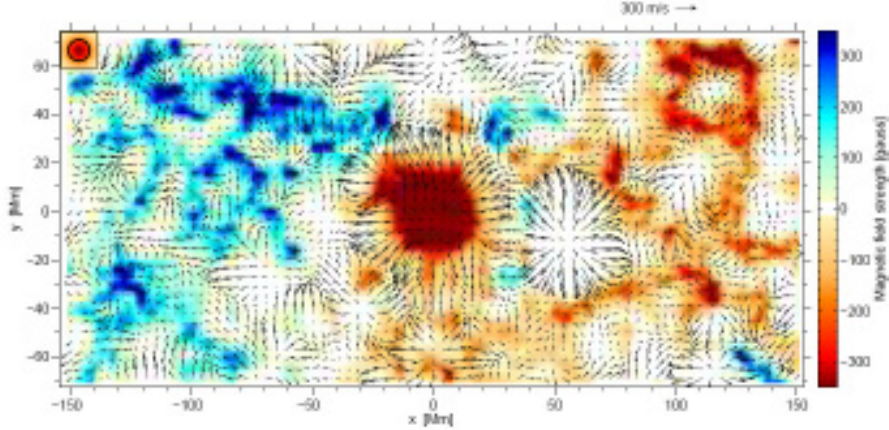}
\end{center}
\caption{
Moat flow around the sunspot in AR~9787 (see {\bf Figure~\ref{fig.observables}}) using the same time-distance inversion as in {\bf Figure~\ref{fig.vz_combo}}{\boldmath $a$}. The background colors show the MDI line-of-sight component of the magnetic field. The depth is $1$~Mm and the observation time is $T=1$~day. The random noise in each horizontal component of the flow is estimated to be $17$~\ms.
Adapted from \citet{Gizon2009} by Jason Jackiewicz.
}
\label{fig.moat}
\end{figure}

Local helioseismology is a useful tool for studying  flows around sunspots
\citep{Duvall1996,Lindsey1996}.
\citet{Gizon2000}  used f-mode time-distance helioseismology to study the
moat flows in the two Mm below the surface.  Comparisons between the
near-surface flows inferred from local helioseismology with direct Doppler
measurements have demonstrated the validity of the near-surface inversions.
{\bf Figure~\ref{fig.moat}} shows the moat flow at a depth of 1~Mm using
a time-distance inversion and all ridges from f through p$_4$ \citep{Gizon2009}.
At this depth, the moat flow has an amplitude of $\sim 250$~\ms, which is consistent with the motion of the small magnetic
features. This outflow is detected in the top 4~Mm.
Such measurements of the subsurface moat flow have been confirmed
by ring-diagram analysis (Moradi et al. 2009, submitted).

The moat flow is believed to be driven by a pressure gradient caused by the blockage of heat transport by sunspots \citep[e.g.][]{Nye1988}. For a more accurate description in terms of the surface cooling that causes the observed flows, see \citet{Spruit1997}. Though much slower, the moat flow appears physically connected with the Evershed flow in the penumbra. This is evident both from the observations by, e.g., \citet{Cabrera2006}
and the realistic numerical simulations of \citet{Heinemann2007} and \citet{Rempel2009}.

\subsection{Absorption of Solar Oscillations}
\label{sec.sunspot_sinks}

The first major discovery made by local helioseismology was that solar oscillations (f and p modes) are absorbed by sunspots \citep{Braun1987}. This discovery was made using Fourier-Hankel analysis, which is based on the decomposition of the wavefield around the sunspot into ingoing and outgoing waves. \citet{Braun1987, Braun1988} found that typical sunspots can absorb up to 50\% of the incoming power.

\citet{Spruit1992} proposed that mode conversion between the oscillations of the quiet Sun and the slow magneto-acoustic waves of the sunspot could explain the observations of wave absorption by  sunspots. Theoretical modeling \citep{Cally1993,Crouch2005} and numerical simulations \citep{Cally1997,Cally2000,Cameron2008} demonstrated that mode conversion is indeed capable of removing a large fraction of the energy from the helioseismic waves incident on a sunspot.  The efficiency of the mode conversion is strongly dependent on the angle of the magnetic field from the vertical, with a maximum in the absorption occurring at an angle of about 30$^\circ$ from the vertical \citep{Cally2003}.  Comparisons between observations and models \citep[e.g.][]{Crouch2005,Cameron2008} show that the explanation of mode conversion is consistent with the observations of the reduction of outgoing wave power.   We note that plage regions are also known to 'absorb' incoming waves.

\subsection{Phase Shifts and Wave-Speed Perturbations}\label{sec.sunspot_structure}

Fourier-Hankel analysis (Section~\ref{sec.hankel}) has demonstrated that outgoing waves from sunspots
have different phases than the corresponding ingoing waves
\citep{Braun1992,Braun1995}.
At fixed radial order, the phase shifts increase
roughly linearly with increasing angular degree. \citet{Fan1995} used a simple
model, in which the sunspot is treated as a local enhancement in the
sound-speed, to suggest that these phase shifts are indicative of a
near-surface change in the wave speed relative to quiet Sun.
\citet{Crouch2005} showed that approximate models of wave propagation in a
model sunspot (embedded magnetic cylinders)
 could simultaneously explain the  absorption and phase shift
measurements. The success of these simple models suggests that the interaction
of solar oscillations with the sunspot magnetic fields may be the essential
physics in understanding both wave absorption (Section~\ref{sec.sunspot_sinks}) and the phase shifts caused by sunspots.

\begin{figure}[t!]
\includegraphics[width=\linewidth]{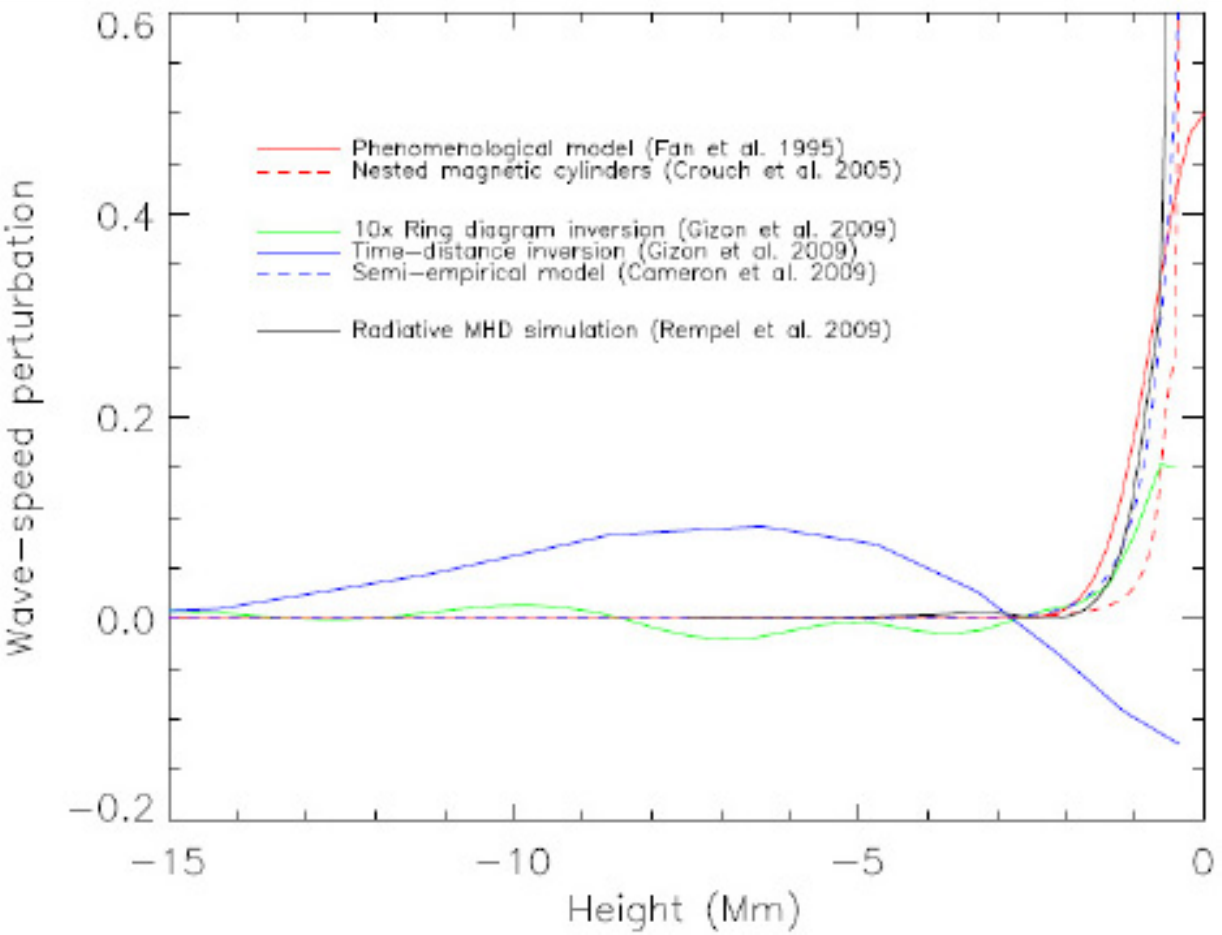}
\caption{
Wave-speed perturbations under sunspots relative to quiet Sun. The perturbations are measured along sunspot axes except for the case of (unresolved) ring-diagram analysis.
The solid red line shows a phenomenological model based on the Fourier-Hankel analysis of the sunspot in active region NOAA~5254 during 27\,--\,30 November 1988 \citep{Fan1995}. The dashed red curve shows the fast wave speed, $c_{\rm f} = (c^2+a^2)^{1/2}$,  in NOAA~5254 from the forward model of \citet{Crouch2005} that consists of nested magnetic cylinders.
The green and solid blue lines give the wave-speed perturbations under the sunspot in active region NOAA~9787 inferred from ring-diagram analysis and time-distance helioseismology using phase-speed filters \citep{Gizon2009}. The same tracked patch (diameter $15^\circ$) was analyzed in both cases.
The sunspot is not spatially resolved in the ring analysis: a factor of ten is applied to improve the comparison.
This active region, which includes a long-lived sunspot and surrounding plage, was observed by SOHO/MDI during 20-27 January 2002.  The two analyses give inconsistent estimates of the subsurface wave-speed perturbations averaged over the $15^\circ$ patch. Possible explanations for this disagreement are described in the text.
The dashed blue line is the fast wave speed of the semi-empirical model of Cameron et al. (2009, submitted), based on the umbral model of \citet{Maltby1986} and discussed in {\bf Figure~\ref{fig.slim}}.
The black curve is the fast wave speed from the radiative MHD numerical simulation of \citet{Rempel2009}.
}
\label{fig.compare_wave_speed}
\end{figure}

Time-distance, holography, and ring-diagrams have all been used to infer
changes in the wave speed in sunspots \citep[e.g.][among a great many others]{Kosovichev1996, Kosovichev2000,Jensen2001,Basu2004,Couvidat2006}. Interpretation of the helioseismic measurements is a rapidly developing topic of current research. Figure~\ref{fig.compare_wave_speed} shows a comparison between time-distance and ring-analysis inversions, forward models based on Fourier-Hankel analysis, and a realistic numerical simulation of a sunspot. As was shown by \citet{Gizon2009} and Moradi et al.\ (2009, submitted), there is not yet agreement among the different analysis methods, especially between the time-distance and ring-diagram results. There are a number of possible explanations for this disagreement. Current inversions for the time-distance and ring-diagram methods use sensitivity functions that do not explicitly include the direct effects of the magnetic field and both assume that wave-speed perturbations are small. The time-distance sensitivity functions may not model the reference power spectrum sufficiently accurately (convective background, mode frequencies, relative mode amplitudes, line widths and asymmetries). Neither method fully accounts for the details of the measurements procedure, especially in the case of time-distance where the effects of the data analysis filtering in Fourier space (e.g., phase-speed filters) are not fully accounted for. Except for the time-distance inversion, all other methods are consistent with an increased wave speed in the top 2~Mm and show wave-speed perturbations with amplitudes less than about 2\% at greater depths.

\begin{figure}[t!]
\includegraphics[width=9.cm]{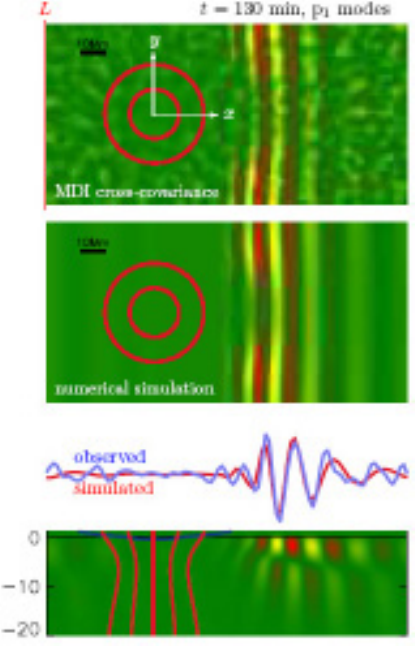}
\caption{\small Sunspot time-distance helioseismology and forward numerical modeling. The top panel shows the observed covariance,  $C(\br, t)=\int dt' \overline{\phi}(t') \phi(\br, t'+t)$, between the MDI Doppler velocity averaged over the red line ($L$) at $x=-40$~Mm, $\overline{\phi}(t')$, and the Doppler velocity delayed by $t=130$~min, $\phi(\br, t'+t)$. The horizontal coordinates $\br=(x,y)$ are centered on the sunspot. The color scale is such that positive values of $C$ are red and negative values are yellow.  The two red circles indicate the boundaries of the umbra and penumbra of the sunspot in Active Region 9787.  The Doppler observations were filtered to select only the p$_1$ acoustic modes. To reduce noise, the cross-covariance was averaged over nine days (20--28 January 2002) and over angles using the azimuthal symmetry of the sunspot. The panel below shows the numerical simulation from {\bf Figure~\ref{fig.slim_vx}}, except that the vertical component of velocity, $v_z$, is now shown. The initial conditions were chosen such that $v_z$ matches the observed cross-covariance in the far field.
The observed $C$ and the simulated $v_z$ are averaged over $-2.5~\mbox{Mm}<y<2.5~\mbox{Mm}$ and plotted as function of $x$. The simulation provides a good match in phase and amplitude with the observations, for this particular model sunspot. The bottom panel shows the simulated $v_z$ in the $x$--$z$ cut through the sunspot. The vertical scale is given in units of Mm and the $a=c$ level is shown by the blue curve.
See {\bf Supplemental Movie 8}.
A similar analysis was performed by \citet{Cameron2008} for f-mode wave packets.
}
\label{fig.slim}
\end{figure}

Direct simulation of wave propagation through sunspot models is useful to test the validity of these models. \citet{Cameron2008} used a three-dimensional MHD code to compute the propagation of f modes through a model sunspot. Here we show a computation for the propagation of a p$_1$ wave packet using the same simulation code. The simulated wavefield, solution to an initial value problem, is compared to the observed cross-covariance in {\bf Figure~\ref{fig.slim}} in order to assess the validity of the underlying sunspot model. Like for the f modes, the comparison with the observations is promising: the amplitude of the transmitted waves is reduced and the waves travel faster in the sunspot than in quiet Sun.

As mentioned in Section~\ref{sec.strong_perturbation}, sunspot magnetic fields strongly affect the nature of the wave properties in the surface layers. In particular, upward propagating high-frequency ($\omega>\omega_{\rm c}$) waves are reflected and refracted at the $a=c$ surface, where MHD mode conversion occurs. \citet{Finsterle2004APJL} used multi-height observations of solar oscillations to map the $a=c$ surface in active regions, called  the `magnetic canopy'. The travel time measured between two observation heights in the solar atmospheres was used to derive the propagation properties of the waves between these two layers. When both heights are above the $a=c$ layer, waves are evanescent and the travel time vanishes. Using combinations of three heights, they find that in sunspots and active regions the canopy deeps below the base of the photosphere by several hundred km, while it is above 1000~km in the quiet Sun.

\section{EXTENDED FLOWS AROUND ACTIVE REGIONS}
\label{sec.SSW}

In this section, we describe flows around large complexes of magnetic activity.
These flows should not be confused with the (smaller scale) moat flow around individual sunspots, which
was discussed in Section~\ref{sec.sunspots}.

\subsection{Surface Inflows, Deeper Outflows}

\begin{figure}[t!]
\includegraphics[width=\linewidth]{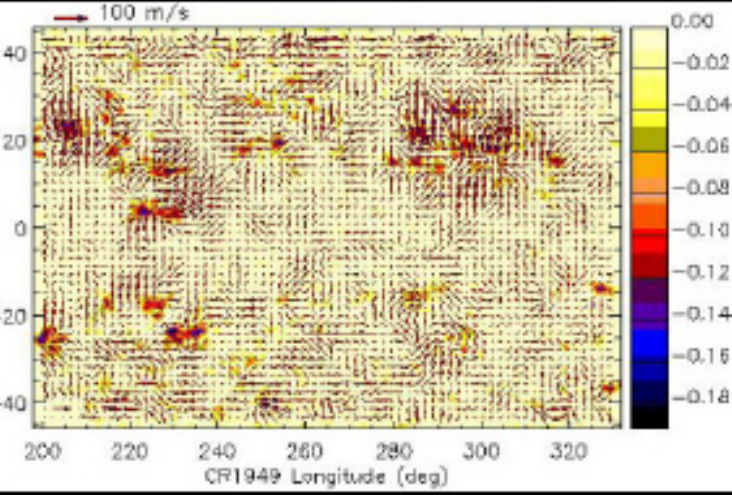}
\caption{
Synoptic map of local horizontal flows in the top 2~Mm below the solar surface,
obtained with f-mode time-distance helioseismology. The horizontal and vertical axes give the longitude and the latitude in heliographic degrees.
The data were averaged in time (7 days) in a frame of reference that co-rotates with the Sun (Carrington rotation rate). The flow maps were further processed to remove the average differential rotation and meridional flow.
The gray scale gives the relative change in f-mode travel times: reduced travel times correlate with magnetic activity. Adapted from \citet{Gizon2001}.
}
\label{fig.ssw_flows_fmode}
\end{figure}

Using f-mode time-distance helioseismology, \citet{Gizon2001} detected weak $\sim 50$~\ms surface flows that converge toward active regions ({\bf Figure~\ref{fig.ssw_flows_fmode}}). These inflows, which exist as far as  $30^\circ$ from the centers of active regions, are also seen in ring diagram analyses \citep[e.g.][]{Haber2001, Haber2004, Komm2007}.
In fact, \citet{Hindman2004} showed that the time-distance and ring-diagram methods
give nearly identical results near the surface.
These near-surface flows also agree reasonably well with the motion of supergranules \citep{Svanda2007}. The converging flows near the surface are accompanied by cyclonic flows with vorticity of order $10^{-7}~{\rm s}^{-1}$ \citep{Komm2007}.

At depths in the range of about 10~Mm to 15~Mm, diverging flows from active regions have been inferred using the time-distance \citep{Zhao2004APJ} and ring-diagram \citep{Haber2004} methods.  These diverging flows typically have amplitudes of order 50~\ms.  \citet{Komm2004} used ring-diagram measurements together with the constraint of mass conservation to infer downward flows of order 1~\ms, at depths less than about 10~Mm, in and around active regions.  Below this depth, the active regions tend to show upflows.

\begin{figure}[t!]
\begin{center}
\includegraphics[width=\linewidth]{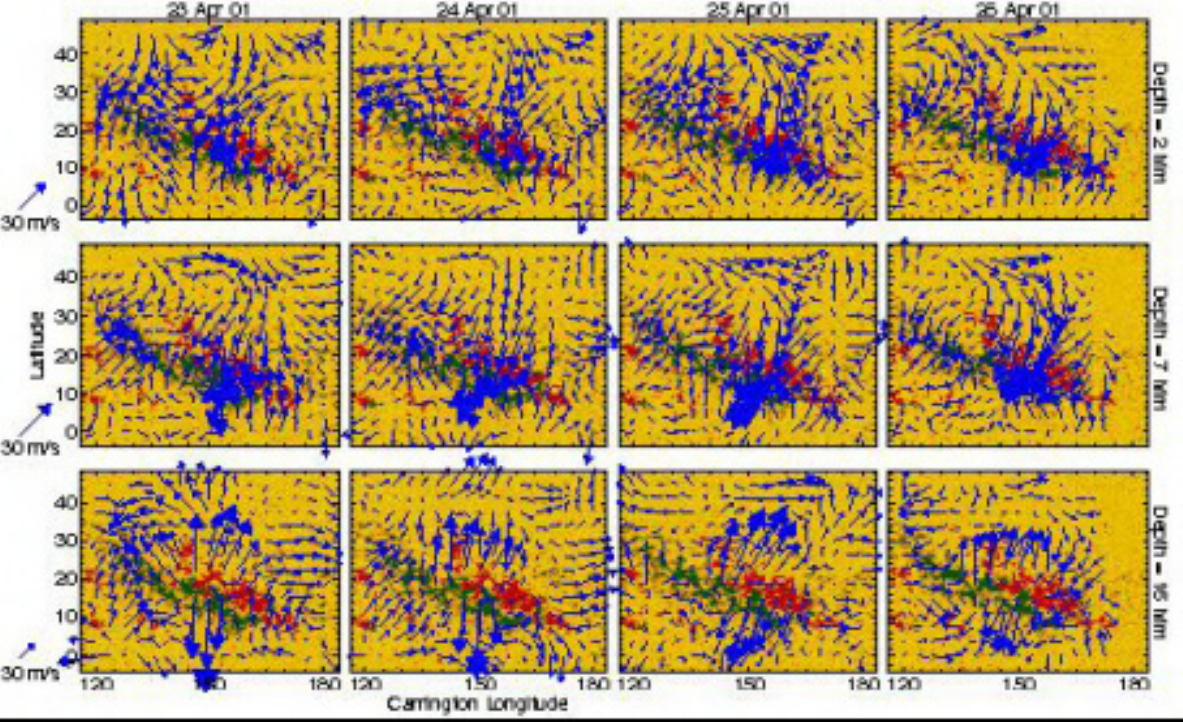}
\end{center}
\caption{
Daily averages of the horizontal flows around active region NOAA 9433 from  23 April until 26 April 2001 (one column for each day) inferred from ring-diagram analysis \citep{Haber2004}. The depths shown are 2~Mm (top row), 7~Mm (middle row) and 14 Mm (lower row). The green and red shades are for the two polarities of the surface magnetic field (MDI magnetograms).  The transition between inflow and outflow occurs near 10 Mm depth.
}
\label{fig.ssw_flows}
\end{figure}

The observations are summarized in {\bf Figure~\ref{fig.ssw_flows}}, which shows the organization of horizontal flows around a particular complex of magnetic activity at three different depths. The flow patterns are consistent from day to day, despite the presence of supergranulation noise.

\subsection{Flows due to Thermal Effects of Magnetic Fields}

A diagnostically important class of flows are those associated with thermal effects due to magnetic fields, such as heating by dissipation of magnetic energy or the enhanced radiative loss of small scale magnetic fields at the surface. On intermediate to large scales  and time scales exceeding the rotation period approximate geostrophic balance holds in the convection zone. Hence flows of the thermal wind type must accompany thermal disturbances on those scales. For example,  the enhanced radiative loss from the small scale magnetic field in active regions has a cooling effect that should drive
an inflow at the surface and a circulation in the cyclonic sense around the active region \citep{Spruit2003}, as in low pressure systems in the Earth's atmosphere.
This may be an explanation for the flows described in the previous section.

\section{GLOBAL SCALES}\label{sec.global_scales}

The dominant global-scale flows in the Sun are the  differential rotation and
the meridional flow.  Helioseismic measurements made over long time
scales (a few solar rotation periods) effectively remove the contribution of
small-scale convective flows and active region flows
and allow high precision studies of these
global-scale flows.  Both rotation and the meridional flow show small
variations with the solar cycle.

\subsection{Differential Rotation}
\label{sec.rot}

The North-South symmetric component of internal differential rotation has been
measured using global helioseismology \citep[cf. review by][]{Thompson2003}.  The solar rotation rate depends strongly on latitude in the convection zone, with the equator rotating more quickly than the poles.
The rotation rate shows only a weak radial shear in the bulk of the convection
zone.  There is strong radial shear in the very near surface layers
\citep[top 35~Mm,][]{Schou1998}, and in the tachocline where the differentially rotating convection zone meets the uniformly rotating radiative zone.  The tachocline plays an important role in most dynamo theories of the solar cycle.

As the differential rotation is well known, it provides an important test of
local helioseismic methods. \citet{Giles1998}
measured rotation with time-distance helioseismology applied to MDI data and found good
qualitative agreement with the results of global
helioseismology. \citet{Basu1999} and \citet{Gonzalez2006} both used
ring-diagram analysis to study the differential rotation in the near-surface
shear layer.  These studies found rotation rates that essentially agreed
with those inferred from global helioseismology in the top 20~Mm (increasing angular velocity with depth), below that depth the measurements showed instrument dependent systematic errors.

\subsection{Meridional Flow}\label{sec.merid}

Doppler measurements \citep[e.g.][]{Hathaway1996} reveal a surface meridional flow with an amplitude of about 20~\ms from the equator to the poles.  It has also been measured by tracking magnetic elements \citep{Komm1993MER}, with essentially the same result.  This surface meridional flow implies a subsurface return (i.e., equatorward) flow.

In flux-transport dynamo models \citep[e.g.][]{Dikpati1999,Charbonneau2005}, %Dikpati1999, Dikpati2006},
the meridional flow is responsible for the redistribution of
flux from the active latitudes to the poles (at the surface) and in some models also
from the poles to equator (at the base of the convection zone).
\citet{Hathaway2003} argued that the equatorward
drift of sunspots during the course of the solar cycle (the butterfly diagram)
implies the existence of a $1.2$~\ms return flow at the bottom of the convection zone.
However \citet{Schuessler2004} argue that the butterfly diagram could
be reproduced by a traditional model of dynamo waves without transport of
magnetic flux by a flow.

In addition to its role in some dynamo theories, the meridional flow is
also thought to play an important role in transporting angular momentum
and thus in maintaining the differential rotation \cite[for a recent review see][]{Miesch2005}.

The meridional flow produces a second-order shift in the frequencies of the
global modes (unlike, for example, the differential rotation which produces a
first-order shift) and is therefore very difficult to measure using
traditional global helioseismology.  The meridional flow does however produce
a first-order change in the eigenfunctions of the modes of the Sun, and thus produces first-order effects in local helioseismic measurements.

\citet{Giles1997} used time-distance helioseismology to obtain the first
detection of the subsurface meridional flow. Imposing a mass conservation
constraint, inversions of the time-distance measurements suggested a
meridional flow that fills the convection zone, is equatorward below about
$0.8 R_\odot$, and has a strength of about 2~\ms at the base of the convection
zone \citep{Giles2000}.
These deep results, however, are not direct measurements.

\begin{figure}[t!]
\includegraphics[width=7.cm]{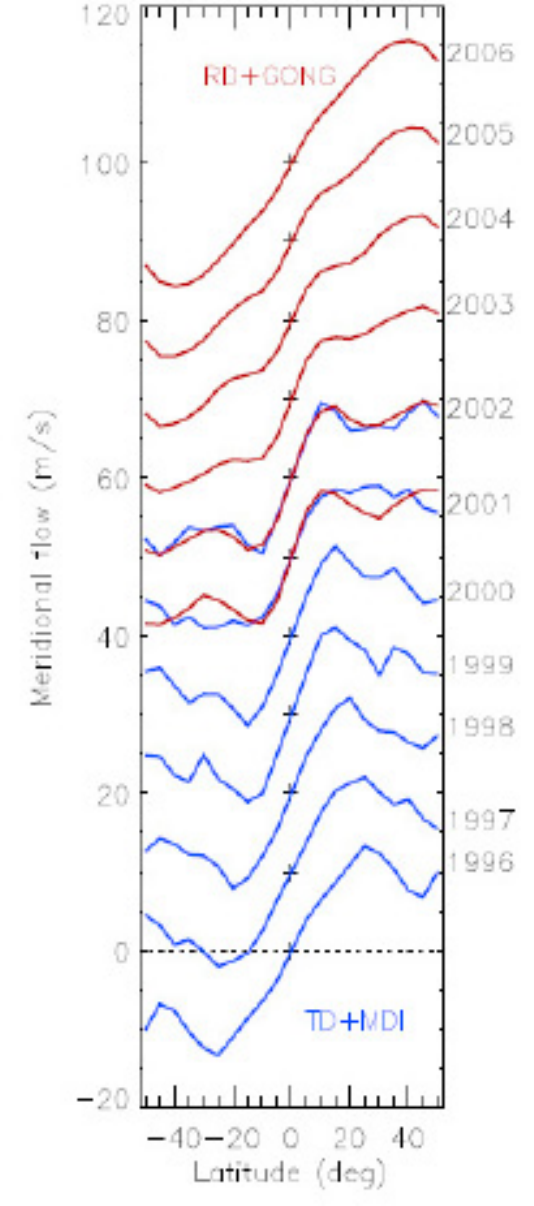}
\caption{Anti-symmetric component of the near-surface meridional flow
  as a function of latitude during 1996\,--\,2006.  Each curve corresponds to a  different year as
  indicated on the right. To improve lisibility, years 1997 and above
are shifted by multiples of 10~\ms.
The blue curves until 2002 show
 the advection of supergranulation as measured by time-distance
 helioseismology \citep{Gizon2008} and
 MDI full-disk data (2\,--\,3 months per year).
The red curves from 2001 are averages (from the surface down to 7 Mm) of the meridional flow
 inferred by ring-diagram analysis and GONG data
\citep{Gonzalez2008SOL}. The ring-diagram values are multiplied by a factor
of $0.8$ to ease the comparison.
Note the local maximum moving towards the equator, from $25^\circ$ in 1996 to
$10^\circ$ in 2006.}
\label{fig.meridional_flow}
\end{figure}

{\bf Figure~\ref{fig.meridional_flow}} shows measurements of the meridional
flow close to the surface, using ring-diagram analysis and (a variant of)
time-distance helioseismology. The maximum amplitude of the meridional flow
is about $15$--$20$~m/s. The eleven years of data show that the meridional
flow varies significantly (up to 50\% of its mean value).
The solar-cycle dependence of the meridional flow will be discussed in the
next section.

As discussed by \citet{Braun2008APJL}, one of the fundamental difficulties in measuring the deep meridional flow is that the noise level, due to the stochastic nature of solar oscillations, is very large compared to the weak signal expected from a flow of a few \ms (for comparison, the sound speed at the base of the convection zone is roughly 230~\kms).

Numerical simulations of convection in rotating shells \citep{Miesch2006} have
roughly reproduced the overall amplitude of the meridional flow, although in
these simulations the meridional flow is highly variable and has a
multiple-cell structure. These simulations rely on the presence of a small latitudinal entropy gradient to establish solar-like differential rotation as suggested by \citet{Rempel2005}.
This gradient is imposed at the base of the convection zone model as an adjustable part of the boundary conditions. The differential rotation in these models is thus what in geophysics is called a 'thermal wind', much like the global atmospheric circulation on earth is due to the pole-equator temperature difference.  In the Sun the cause for such latitudinal entropy variation is not quite clear, however. Independent of this unsolved question, the fact that the simulations so far appear unable to reproduce the solar differential rotation without an imposed entropy gradient  is significant. It implies that models of the Sun's differential rotation based on $\Lambda$-effect or anisotropic turbulence formalisms so far are not substantiated by numerical simulations.

\subsection{Solar-Cycle Variations}\label{sec.torsional_oscillations}

\begin{figure}[t!]
\includegraphics[width=\linewidth]{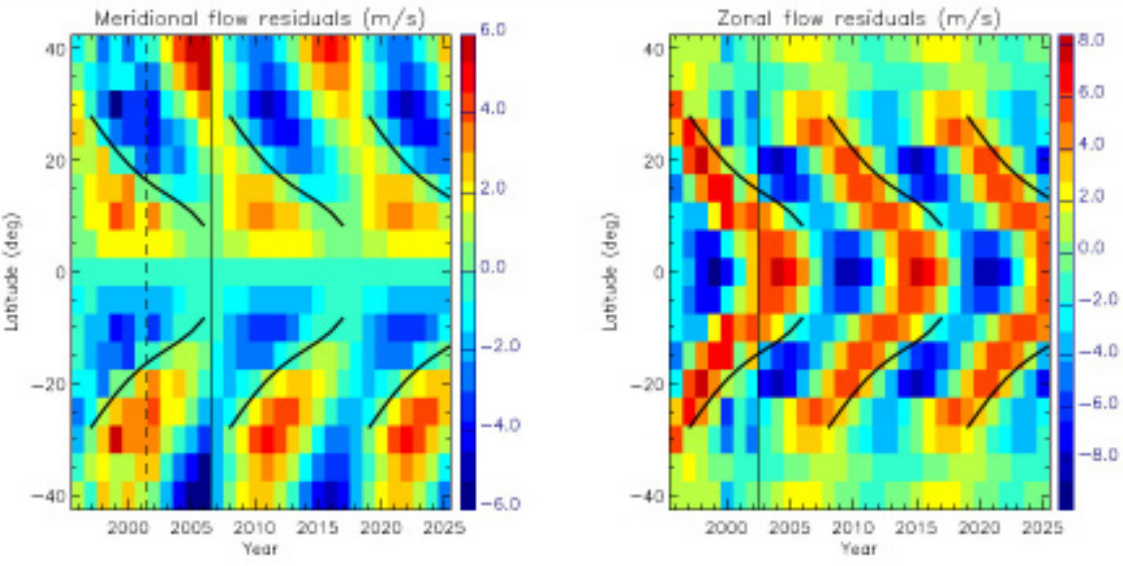}
\caption{Solar cycle variations of the meridional and zonal flows
in the near-surface layers.
({\it Top}) Time residuals of the meridional flow
after subtraction of the average meridional flow for
 1996--2006 (from {\bf Figure~\ref{fig.meridional_flow}}).
The color scale is in units of \ms.
The first eleven years (the observations) are
extrapolated into the future by fitting the
eleven-year periodic component.
The thick black curves represent the mean latitude of activity
estimated from Mount Wilson magnetograms.
({\it Bottom}) Time residuals of the zonal flows
after subtraction of the mean rotational velocity
for the period 1996--2002, followed by the eleven-year periodic
component. Flows are deduced from the
advection of the supergranulation in MDI time-distance
divergence maps \citep{Gizon2008}.
The color scale is in units of \ms.
}
\label{fig.residuals_surface}
\end{figure}

The time varying component of rotation shows bands of slower and faster rotation ($\pm 10$~\ms) that migrate in latitude with the phase of the solar cycle. This pattern, called torsional oscillations, was first seen in direct Doppler measurements \citep{Howard1980}.
The torsional oscillations have two branches.  At latitudes less than about 45$^\circ$,
the bands of increased and decreased rotation move towards the equator at the same rate as the active latitudes, with the active latitudes located on the
poleward side of the fast band.  At high latitudes, the bands move towards the poles.
Global helioseismology has shown that the torsional oscillations have their maximum amplitude close to the surface, but extend throughout much of the convection zone \citep[e.g.][]{Vorontsov2002}. The torsional oscillations tend to be roughly uniform along contours of
constant rotation rate \citep[e.g.][]{Howe2005}. As reviewed by \citet{Gizon2005}, local helioseismology has been used to confirm many of these results for the shallow component of the torsional oscillations.
{\bf Figure~\ref{fig.residuals_surface}} shows the time residuals of the zonal flows near the surface  (time-distance helioseismology).

In addition, local helioseismology has shown that there are fluctuations,
with amplitudes of $\pm 5$~\ms,  in the meridional flow that are associated
with the butterfly diagram \citep[this pattern was seen in direct Doppler
measurements by][]{Ulrich1988}.  As shown in {\bf Figure~\ref{fig.residuals_surface}},
near the surface these local fluctuations
correspond to flows converging towards the active
latitudes \citep[e.g.][]{Gizon2008,Gonzalez2008SOL}.
The exact depth where these flows change sign is not well known, but at depths of roughly 50~Mm, the
component of the meridional flow associated with the torsional oscillations
converges towards the active latitudes  \citep[e.g.][]{Chou2001APJL,Beck2002}.
The contribution of flows around individual active regions to the torsional
oscillations and associated meridional flows will be discussed in Section~\ref{sec.SSW}.

\citet{Schuessler1981} and \citet{Yoshimura1981} suggested that the torsional oscillations may be caused by the Lorentz force associated with migrating dynamo waves.
A turbulent mean field dynamo model by \citet{Covas2000}, fitted to a butterfly diagram of the solar cycle, shows a Lorentz force-induced torsional oscillation pattern at the surface resembling the observations.
As in other Lorentz-force models, however, its amplitude increases strongly with depth, in contrast with the helioseismic measurements.
\citet{Kitchatinov1999}, building on work by \citet{Kueker1996}, suggested that the torsional oscillations result  from the effect of the magnetic field on the convective transport of angular momentum. Another suggested explanation is the reduction of turbulent viscosity in active regions \citep{Petrovay2002}.

\citet{Spruit2003} suggested that the torsional oscillations may be a result of geostrophic flows set up by enhanced surface cooling in regions of magnetic activity. Since the driving force in this explanation is at the surface, the velocity signal produced decreases with depth as observed. \citet{Rempel2007}  argued that such a thermal forcing, rather than mechanical forcing as in the Lorentz-force based models, is required to explain the observed deviation of the low-latitude torsional oscillations from a Taylor-Proudman state (zonal velocity constant on cylinders).   Similarly, \citet{Gizon2008} suggested that the only current model for the low-latitude branch of the torsional oscillations and associated meridional flows that is qualitatively consistent with the observations is the enhanced cooling model of \citet{Spruit2003}.  One complication for models that invoke thermal forcing at the surface is to explain the presence of the torsional oscillations at solar minimum \citep{Gizon2008}.
It should be noted that the model of \citet{Rempel2007} does not require enhanced thermal forcing at high latitudes ($>50^\circ$) to explain the poleward-propagating branch of the torsional oscillations. The two branches of the zonal torsional oscillations may have different physical origins.

\subsection{Contribution of Active Region Flows to Longitudinal Averages}

An interesting question is whether the localized flows around active regions (Section~\ref{sec.SSW}) contribute significantly to the solar-cycle variation of the longitudinal averages of the differential rotation and the meridional flow. The inflows/outflows around active regions could affect the average meridional circulation around the mean latitude of activity, while the vortical component of the flows could affect the average zonal flows.

In order to study this question, \citet{Gizon2003PHD} selected all regions within $5^\circ$ of all locations with strong magnetic field and excluded these regions from the longitudinal averages of the flows. The zonal flows are essentially unaffected (except for the fact that active regions rotate a little more rapidly than quiet Sun): localized cyclonic flows around large active regions do not provide a sufficient explanation for the torsional oscillations. This is not particularly surprising since torsional oscillations exist at solar minimum, in the absence of active regions. The torsional oscillations model of \citet{Spruit2003} would have to rely on the thermal disturbances caused by diffuse small-scale magnetic fields, not localized active regions. On the other hand, \citet{Gizon2003PHD} finds that the inflows around active regions appear to be largely responsible for the near-surface solar-cycle dependence of the meridional flow, at the level of a few \ms. This conclusion has been challenged however by \citet{Gonzalez2008SOL}, which indicates that the answer depends sensitively on the selection of the regions of activity that are removed from the longitudinal averages.

\section{FARSIDE IMAGING}
\label{sec.farside}

\citet{Lindsey2000SCI} introduced the concept of farside imaging,
in which observations of the solar oscillations made on the visible disk
are used to infer the presence of active regions on the farside of the Sun.
Farside imaging has been achieved using both holography-based methods
\citep[e.g.][]{Lindsey2000SCI} and time-distance helioseismology \citep{Zhao2007}.
\citet{Hartlep2008} have successfully tested farside time-distance helioseismology with numerical simulations.

\begin{figure}[h!]
\includegraphics[width=8.cm]{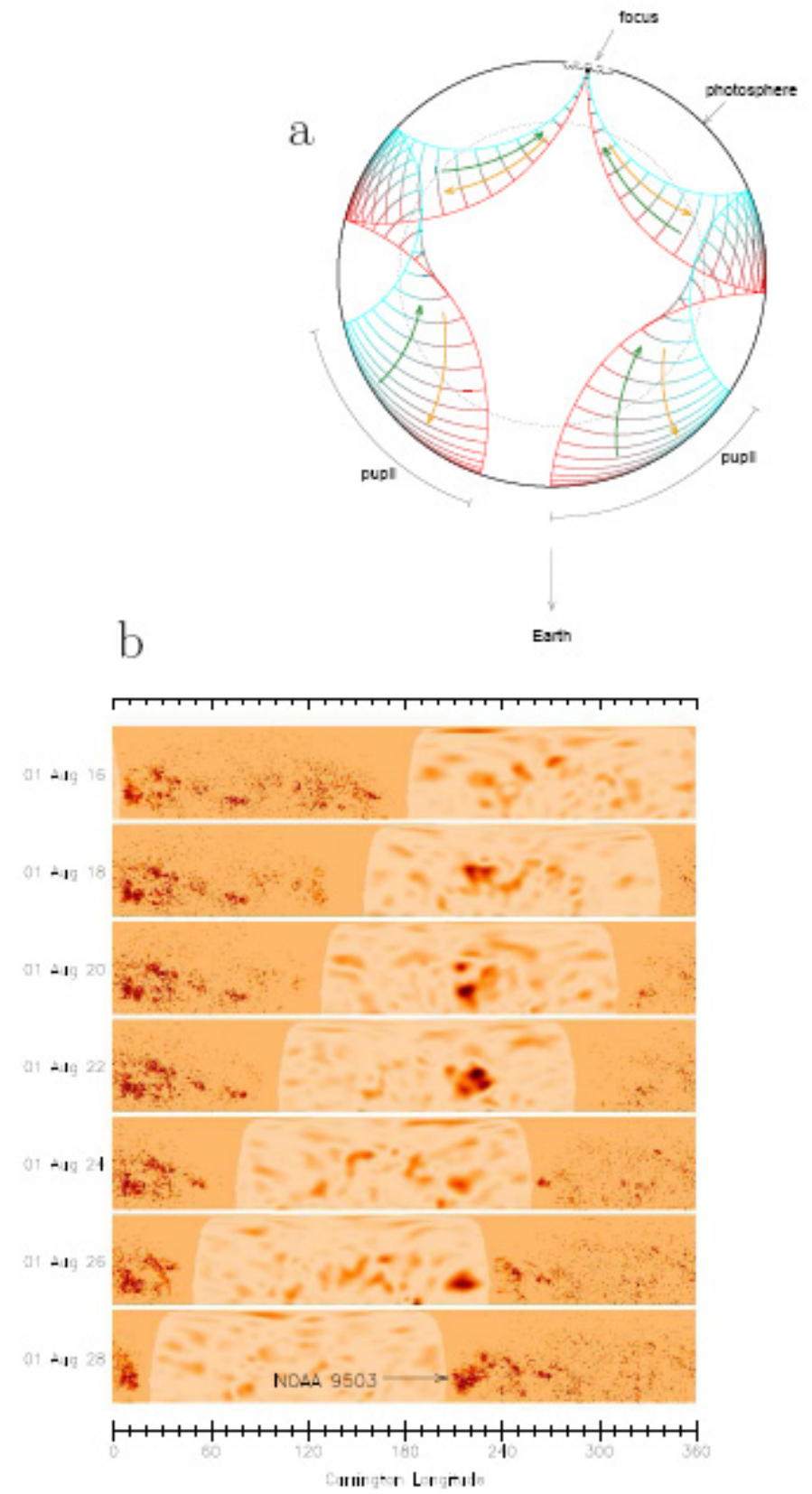}
\caption{
Farside imaging. ({\it a}) Ray-path diagram for 2+2 skip farside imaging \citep{Lindsey2000SCI}. Waves seen leaving (arriving) at the pupils on the visible surface of the Sun travel in the direction of the yellow (green) arrows.  The waves skip from the solar surface once before, and once after, reaching the focal point on the farside of the Sun.  Active regions located at the focal point induce small phase shifts into the waves. \label{fig.farside_rays}
({\it b}) GONG farside images (lighter yellow background) combined with magnetograms
of the front side (darker yellow background) covering a 12~day period each. Each full-Sun map is plotted as a function of Carrington longitude (latitude in a co-rotating frame) and latitude.  The active region NOAA~9503 (August 2001) is seen to form on the farside of the Sun and then rotate onto the visible disk.  Courtesy of Charles Lindsey.}
\label{fig.farside_maps}
\end{figure}

The conceptual ray geometry for farside imaging is shown in {\bf Figure~\ref{fig.farside_rays}}{\boldmath $a$}.
In the 2+2 skip geometry, wave packets leave the visible surface, make two skips in the solar interior
(this involves one reflection from the surface), interact with possible surface magnetic activity
on the farside, make two more skips, and are then seen again on the front side. The total travel time
of the wave packet is sensitive to the presence of large active regions on the farside.
Travel-time reductions of up to ten seconds are typically observed for large active regions.
By moving the farside target location, a map of the farside magnetic activity can be constructed.
The 2+2 skip geometry is suitable for mapping regions that are not too far from the antipode of
the center of the visible disk. In order to complete the farside maps, other skip geometries
are required \citep{Braun2001}. The 1+3 skip geometry targets active regions closer to the limb.
In this  geometry, the wave packets only skip once before they reach the target location,
and then skip three times before they are observed again.
Oslund \& Scherrer (2006, unpublished)
combined farside maps from the 2+2 and 1+3 skip geometries to make maps of the entire farside of the Sun. These farside maps are produced daily using the MDI/SOHO data and are available
online (see {\bf Related Resources}).

{\bf Figure~\ref{fig.farside_maps}}{\boldmath $b$} and {\bf Supplemental Movies 9 and 10} and shows a sequence of farside maps that show a large active region moving across the farside and front-side of the Sun. This active region, NOAA~9503, was seen to form on
the farside of the Sun before appearing on the visible disk about 12~days later.
Magnetic maps of the farside provide up to two weeks of
advance warning before large active regions rotate onto the visible disk,
and thus are expected to play an important role in predicting space weather.

In order to interpret the farside maps in terms of physical variables, such as the total unsigned magnetic flux, \citet{Gonzalez2007} have proposed to calibrate the farside images using long lived active regions that are seen before and after they appear on the farside. Future Sun-orbiting spacecraft carrying solar and magnetic imagers (e.g., Solar Orbiter) will provide enhanced opportunities for detailed farside calibration.

\clearpage

\section{FLARE-EXCITED WAVES}
\label{sec.flares}
\begin{figure}[t!]
\includegraphics[width=14.cm]{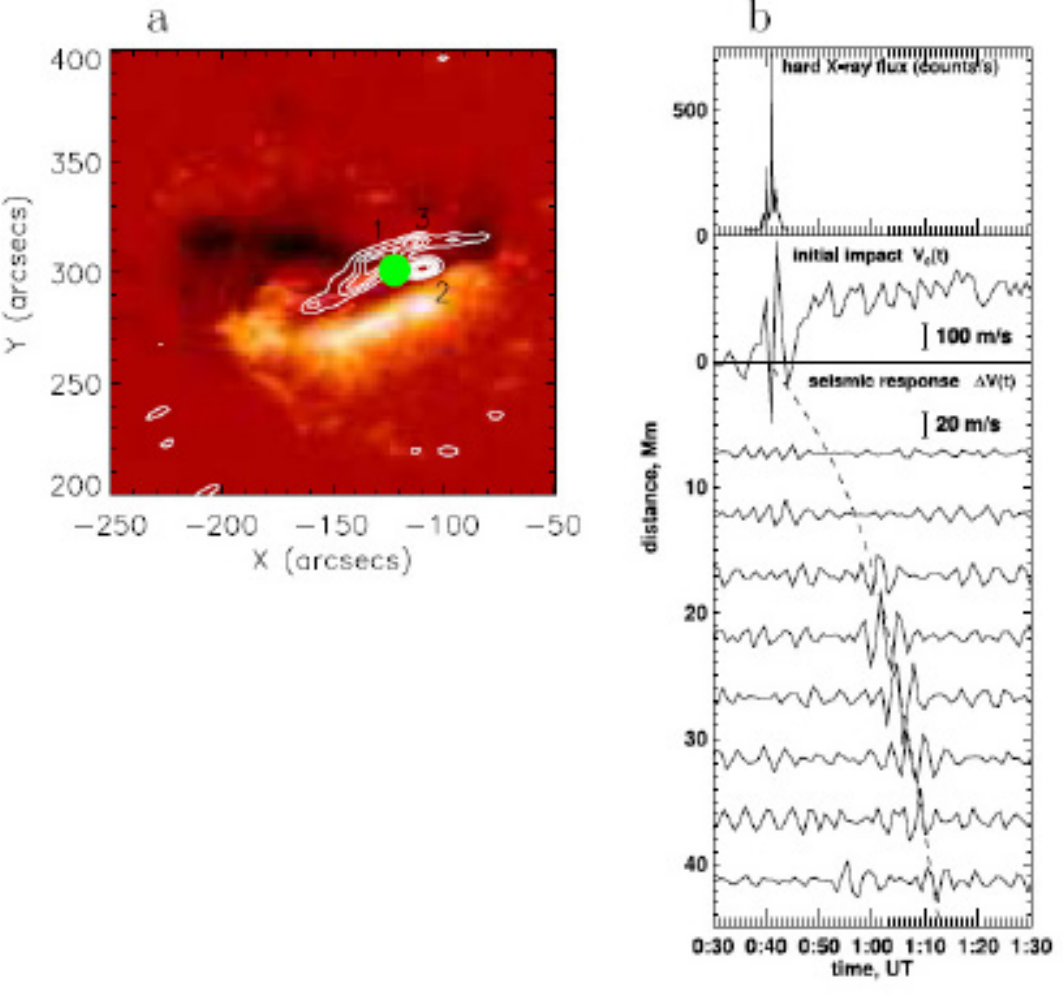}
\caption{
Flare-excited waves in Active Region NOAA 10720 on 15 January 2005. ({\it a}) MDI/SOHO magnetogram (color background) with RHESSI X-ray emission averaged over the period 00:41:33--00:42:34 UT (12--25 keV contours at 10, 20, 30, 40, 50, 60, and 80 per cent of the maximum flux). Three hard X-ray sources are observed. The green dot shows the location of the helioseismic source \citep{Moradi2007}. Courtesy of Alina Donea and Hamed Moradi. ({\it b}) Sequence of events. High-energy electrons accelerated in the solar flare interact with the lower atmosphere, producing hard X-ray emission (observed by RHESSY) and shocks leading to an initial hydrodynamic impact in the photosphere (observed by SOHO/MDI). Raw SOHO/MDI Dopplergrams reveal an expanding seismic wave about 20~min after the initial impact. The dashed curve shows a theoretical time-distance curve for helioseismic waves. Figure and caption from \citet{Kosovichev2006SOL}. With kind permission of Springer Science and Business Media.
}
\label{fig.flares}
\end{figure}

The excitation of solar oscillations by a flare was first observed by
\citet{Kosovichev1998Nature} using
MDI data.  Since then, many other examples have been found
\citep[e.g.][]{Donea2006}.    {\bf Figure~\ref{fig.flares}} shows a summary of
some observations of the waves generated by the flare of 15 January
2005.  In this example, the location of the wave source (as estimated
from helioseismic holography)
is seen to nearly coincide with the hard x-ray emission as
seen by the RHESSI spacecraft.    The seismic waves
are first seen about twenty minutes after the hard x-ray emission and
propagate outwards according to the time-distance relation
({\bf Figure~\ref{fig.ray_paths}}{\boldmath $a$}).
In addition to exciting local waves,
solar flares are also observed to put energy into the the very low
degree global modes \citet{Karoff2008}.

The details of the physical mechanism responsible for the wave
excitation are not clear.
Two main suggestions have been high-energy electrons \citep{Kosovichev2007} and the Lorentz forces \citep{Hudson2008} due to the
reconfiguration of the magnetic field during the flare.
\citet{Lindsey2008} discuss the competing mechanisms in detail.
Additional observations and modeling efforts are needed in order to test these
proposals.

\section{FUTURE OBSERVATIONS}
\label{sec.future}
\subsection{Solar Dynamics Observatory}
The Helioseismic and Magnetic Imager (HMI) is designed
to deliver ideal data for local helioseismology. HMI is one of several
instruments onboard NASA's Solar Dynamics Observatory (SDO) to be flown this
year in a geosynchronous orbit.
HMI will transmit 4086$\times$4096 pixel Doppler images
of the Sun at the cadence of one image every 50~s or better.
It will combine high
spatial resolution (1~arcsec) and full spatial coverage, with a very high duty cycle over a nominal mission duration of five years. This combination will make possible the local helioseismic analysis of regions closer to the limb (less foreshortening), in order to study higher solar latitudes and the evolution of magnetic active regions as they rotate from limb to limb across the solar disk.
In addition to Dopplergrams, HMI will provide images of the three components
of the vector magnetic field, providing important
information for the interpretation of helioseismic data.
A stated goal of HMI/SDO is the
subsurface detection of the magnetic field before it emerges at the
surface, leading to reliable predictive capability
\citep{Kosovichev2007HMI}.
In combination with observations from the Atmospheric Imaging Assembly (AIA),
a set of four SDO telescopes designed to provide an unprecedented view of the
lower corona, HMI and local helioseismology will help
establish relationships between the internal structure
and dynamics of the Sun
and the various components of magnetic activity in the solar atmosphere.
SDO instruments will generate a total data flow of about $1.5$~TB/day, which
represents a real challenge for the ground segment
in terms of  data storage, processing, and analysis.

\begin{figure}[t!]
\centering
\includegraphics[width=7.cm]{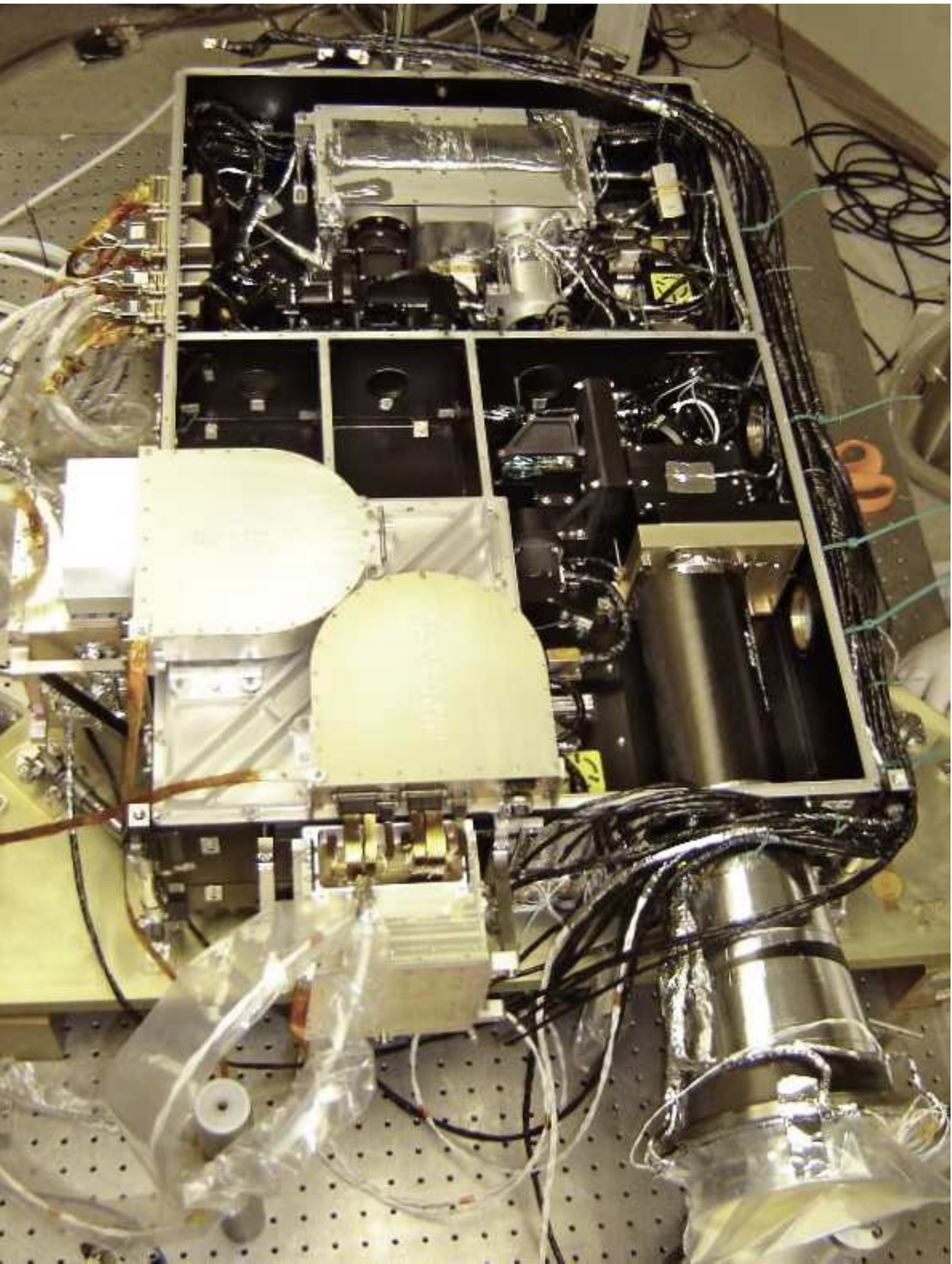}
\caption{The HMI instrument to be flown in 2010 onboard  NASA's Solar Dynamics Observatory.  Courtesy of Philip Scherrer.}
\label{fig.hmi}
\end{figure}

The instrumental design of HMI is similar to MDI's, except for the choice
of the absorption line. Using a combined Lyot-Michelson filter system,
HMI will take five filtergrams across the Fe~I line at 6173~\AA,
separated by $69$~m\AA\ \citep{Borrero2007}.
This line has a Lande factor $g=2.5$
and therefore is better suited for the measurement of the
vector magnetic field (than e.g. the Ni~6768 line).
A picture of the HMI flight model is shown in {\bf  Figure~\ref{fig.hmi}}.
The HMI instrument was delivered in November 2007 and has been integrated onto
the SDO spacecraft. At the time of writing, the launch of SDO is
scheduled for February 2010 from Cape Canaveral.

\subsection{Solar Orbiter}
Solar Orbiter is the next solar physics mission of the European Space Agency
(ESA) and a logical step after SOHO. The target launch date is 2015.
Solar Orbiter will use multiple gravity
assist manoeuvres at Venus and the Earth such that the inclination of the
orbit to the ecliptic will incrementally increase during the course of
the mission (about 10~years) to reach  heliographic latitudes of at
least $30^\circ$.  The elliptical
orbit will have a minimum perihelion distance of $0.22$~AU.
The scientific payload will include a remote sensing package that will
deliver $0.5$~arcsec pixel images of the solar photosphere (intensity, Doppler
velocity, and magnetic field).

While the exact details of the orbit (and observation windows)  are still being discussed,
it is clear that Solar Orbiter will offer unique opportunities
for helioseismology \citep{Woch2007}.
First, it will be possible  to
study the subsurface flows and structure
in the polar regions,
which is not possible today
and is important to understand the solar cycle.
Second, Solar Orbiter will enable us to test and apply the concept of
stereoscopic helioseismology.
Stereoscopic helioseismology
combines observations from different vantage points.
Solar Orbiter's orbit is particularly interesting
as it will offer a large range of spacecraft-Sun-Earth angles.
With observations from two widely different viewing angles (Solar Orbiter
and another Earth or near-Earth experiment), it becomes possible
to consider acoustic ray paths with very large separation distances
(see {\bf Figure~\ref{fig.ray_paths}}{\boldmath $b$}).
This is useful in local helioseismology
 to probe structures deep into the Sun, and, in particular,
at the bottom of the convection zone.

\section{SUMMARY AND OUTLOOK}

Local helioseismology exploits the information contained in the local dispersion relation of the acoustic and surface-gravity waves (ring-diagram analysis) and in the correlations of the random wave field (time-distance helioseismology and related methods) in order to study the subsurface structure and dynamics of the Sun in three dimensions. The high-quality observations from the GONG network and the SOHO satellite have made possible the study of the properties of the upper layers of the convection zone and their variations with the solar cycle.

Local helioseismology has not reached maturity and there are many open questions about data analysis methods and interpretation. The observational results which, in our view, are the most robust and physically sensible are sketched in {\bf Figure~\ref{fig.sketch}} and listed in the {\bf Summary Points} below. Local helioseismology measures effects that are subtle, such as velocities of only a few \ms. In addition to approximations in the data interpretation, it is important to keep in mind that several sources instrumental errors can affect the measurements, e.g., plate scale errors and optical distortion \citep{Korzennik2004} or uncertainties in the orientation of the image \citep{Giles2000}.

\begin{figure}[t!]
\begin{picture}(100,100)
%f25
%\put(0,0){\includegraphics[width=8.cm]{figures/fig5.eps}}
%\put(0,50){\includegraphics[width=4.cm]{figures/matthias_power.eps}}
\end{picture}
\caption{
%Summary sketch (drawing) of the results listed in the Summary Points, especially flow measurements on various spatial scales. This plot is not easy and will require some interaction with our production editor...
}
\label{fig.sketch}
\end{figure}

An important challenge for future work in local helioseismology is to detect signatures of magnetic fields at the base of the convection zone, where the field is expected to be amplified by differential rotation and stored until erupting to the surface as active regions. Direct detection through their effect on wave propagation properties is  unlikely. Because of the high pressure at the base of the convection zone, the contrast in propagation speed is very much lower than in surface structures like sunspots, even at the inferred field strengths of $\sim 10^5$G.
More promising is the prospect of detecting systematic flows that might be associated with magnetic structures at the base of the convection zone. Easiest to detect would be azimuthal flows (variations in rotation rate), such as have already been reported on the time scale of the solar cycle.
Even if the sensitivity of helioseismic methods turns out insufficient to detect such deeply seated structures, it may well be sufficient to rule out certain less-preferred classes of models for the solar cycle, such as convective dynamo models acting throughout the convection zone or in a shallow surface layer. An important class of flows would be the geostrophic flows caused by thermal effects of magnetic fields (see Section~\ref{sec.SSW}). Such disturbances are much easier to detect through their thermal winds than directly by their temperature contrast. They might be turned into a diagnostic of magnetic fields in deeper layers that can be probed with helioseismology.

The availability of powerful computers provides exciting opportunities to devise, validate, and optimize improved methods of local helioseismology. Exploring these possibilities will be key to taking full advantage of the observations of solar oscillations.
In Sections~\ref{sec.strong_perturbation} and~\ref{sec.sunspots}, we have shown examples of the usefulness of numerical simulations of wave propagation through prescribed reference sunspot models.
Simulations of wave propagation in spherical geometry \citep[e.g.][]{Hanasoge2006} have been used in time-distance studies of the deep convection zone \citep[e.g.][]{Zhao2009} and to validate far-side imaging \citep{Hartlep2008}.

\begin{figure}[t!]
\includegraphics[width=\linewidth]{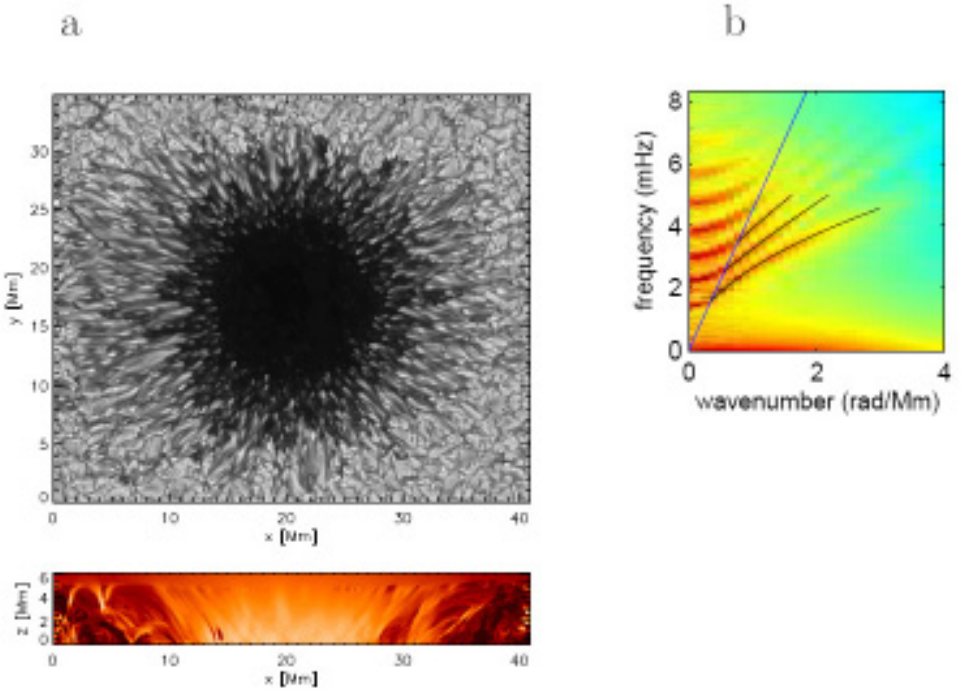}
\caption{Radiative MHD simulation of a sunspot by \citet{Rempel2009}.
({\it a}) Bolometric intensity (black and white) and subsurface magnetic
field strength on a vertical cut through the center of the sunspot (in the range $0$\,--\,$8$~kG).
See {\bf Supplemental Movie 11}.
({\it b}) Power spectrum of the surface oscillations in the simulation.
The blue line is the phase speed at the bottom of the box, above which the model is not realistic.
}
\label{fig.matthiasspot}
\end{figure}

It is now becoming possible to simulate the near surface layers of the Sun, including pores and sunspots, by numerically solving the radiative MHD equations \citep{Rempel2009}. {\bf Figure~\ref{fig.matthiasspot}}{\boldmath $a$} shows a snapshot of the intensity and the magnetic field for a sunspot simulation. In this simulation, the solar oscillations are naturally excited by the convection (see {\bf Figure~\ref{fig.matthiasspot}}{\boldmath $b$}). This type of simulation provides a means for computing realistic time series of Dopplergrams, which can be used as input to all the methods of local helioseismology. With this type of data set, it will be possible to resolve some of the outstanding issues, for example regarding sunspot subsurface structure (Section~\ref{sec.sunspot_structure}).
The recently achieved convergence of observations and realistic 3D radiative MHD simulations of sunspots can count as a major success story in solar physics. It adds confidence in our numerical methods and in our understanding of the physics of solar magnetic activity.

There are many complications in local helioseismology that have not been studied in detail, e.g. instrumental artifacts (point spread function, astigmatism, plate scale), interpretation of the observable (e.g., filtergrams used to construct Dopplergrams) in terms of physical conditions in the solar atmosphere, center-to-limb effects such as foreshortening, and light-of-sight projection of the solar velocity.
 Other complications are related to the physics of wave propagation, e.g. surface magnetic effects, scattering by time-varying heterogeneities (turbulence), multiple scattering, and physical description of wave excitation and attenuation. Understanding and, in some cases, correcting for these issues is needed to apply local helioseismology to challenging problems: deep meridional circulation \citep{Braun2008APJL}, detecting subsurface emerging active regions, high latitudes, statistical description of turbulent flows (e.g. Reynolds stresses), etc. In addition, inferring small amplitude perturbations in the solar interior may require may years of observations and/or appropriate spatial/temporal averaging to optimize signal-to-noise ratio.

Finally, it is worth exploring  the many connections between the results of local helioseismology and global-mode helioseismology: for example, the contribution of active regions to the temporal variations of low-degree mode frequencies, comparisons of rotation measurements (e.g., $1.3$-year tachocline oscillations), deep sound speed anomalies \citep{Zhao2009}, and seismic radii \citep{Kholikov2008,Gonzalez2009}. In principle, local helioseismology should help provide improve surface boundary conditions for global-mode inversions.

\subsection*{SUMMARY POINTS}
\begin{enumerate}
\item
Local helioseismology shows that supergranules are characterized by $\sim 200$~\ms
horizontal outflows and $\sim20$~\ms upflows near the surface.
Magnetic field concentrations are observed at the boundaries of supergranules and
the inclined field provides portals through which low-frequency
waves propagate into the chromosphere.
The correlation between the horizontal divergence of the flow and
the vertical component of vorticity has been measured as a function of latitude:
cyclonic convection is explained by the effect of the Coriolis force.
The pattern of supergranulation has (unexplained) wave-like properties.
\item
The amplitudes, phases, and frequencies of the solar waves are strongly
affected by sunspots. Sunspots ``absorb'' a fraction of the
ingoing waves as they partially convert into downward propagating
slow MHD waves.
Sunspots are surrounded by a horizontal outflow (several hundred \ms)
in an annular region extending as far as twice the penumbral radius.
This moat flow, which persists at least in the top 4~Mm,
is consistent with direct observations of the solar surface.
Little is known about the subsurface magnetic and thermal structure of sunspots.
Forward modeling of the helioseismic wave field requires a surface field of several kG.
Multi-height observations of solar oscillations have been used to map the sunspot magnetic canopy.
\item
Local helioseismology has confirmed the latitudinal differential rotation and the increase of rotation with depth in the top $\sim 35$~Mm of the convection zone (near-surface shear layer). Flows in meridional planes have been measured by local helioseismology in the top $\sim 50$~Mm. For latitudes less than $45^\circ$, the longitudinal component of the flow is poleward, with a maximum amplitude of 15~\ms. It is not clear whether the meridional flow can be detected reliably deeper or at higher latitudes.
\item
The solar-cycle variation of rotation has been confirmed: bands of faster and
slower rotation ($\pm 10$~\ms)  migrate in latitude with magnetic activity.
In addition, local helioseismology has revealed that the longitudinal-average of the
meridional flow also varies with the solar cycle ($\pm 5$~\ms), i.e. by a
significant fraction of its mean value. Near the surface, the time residuals
are consistent with a North-South inflow around the mean latitude of
activity. At a depth of $50$~Mm, the residuals are consistent with a small outflow.
\item
On intermediate scales ($\sim 20^\circ$) weak horizontal inflows ($\sim 50$~\ms)
have been detected around complexes of magnetic activity, near the surface.
If confirmed, these flows
may explain the time evolution of the longitudinal average of the meridional flow.
At greater depths ($>10$~Mm) the horizontal flows appear to switch sign
and diverge from centers of magnetic activity ($\sim 50$~\ms).
In addition, the surface inflows are associated with cyclonic vorticity.
\item
Farside helioseismology works. Large active regions can be detected on the
invisible hemisphere of the Sun, thus providing advanced warning of energetic particle events, days before they occur on the front side.
\end{enumerate}

\subsection*{FUTURE ISSUES}
\begin{enumerate}

\item The most pressing issue in local helioseismology is how to interpret magnetic effects, which requires new methods of analysis. This is illustrated by the fact that the standard methods of analysis yield conflicting inferences regarding sunspot structure and dynamics (see e.g. {\bf Figure~\ref{fig.compare_wave_speed}}). The way forward is to develop methods that incorporate appropriate physical models of the interaction of waves with strong magnetic fields near the surface. Surface magnetic effects must be accounted for before we can detect and study the magnetic field below the photosphere.

%\item The availability of powerful computers provides exciting opportunities to devise, validate, and optimize improved methods of local helioseismology (see examples in {\bf Figures~\ref{fig.slim_vx} and~\ref{fig.matthias}}). Exploring these possibilities will be key to taking full advantage of the observations of solar oscillations.

 \item Instrumental artifacts often dominate realization noise and hamper the study of weak perturbations in the Sun. Ever-improving instrumentation is essential to pushing the limits of local helioseismology, especially to probe the deepest layers of the convection zone and the high-latitude meridional flow. The SDO/HMI instrument---expected to be launched in 2010---represents an important technological step towards improved observations.

% \item How does it all fit together? What consistent physical picture can account for the many discoveries of local helioseismology --  dynamics of near-surface shear layer, active region flows, subsurface meridional flow, solar-cycle variation of large-scale flows, coriolis effects of flows. The dynamo? Be careful, in the best case scenario, need 10 years to, perhaps, measure meriodional flow at the base of the convection zone \citep{Braun2008APJL} Henk, please take care of this one.

\item
Helioseismology has benefited from methods developed for the seismology of the Earth: normal mode theory, travel-time sensitivity kernels, interpretation of the cross-covariance, inverse methods, etc. We expect that local helioseismology will continue to learn from advances in Earth seismology:
notable progress has been made on numerical simulations of wave propagation, the computation of travel time sensitivity kernels using numerical methods, and non-linear inversions of travel times \citep[various aspects of modern seismology are discussed by, e.g.,][]{Tape2009}.

\end{enumerate}

\section*{ABBREVIATIONS/ACRONYMS}
\begin{enumerate}
\item
GONG: Global Oscillation Network Group
\item
SOHO/MDI: Solar and Heliospheric Observatory/Michelson Doppler Imager
\item
SDO/HMI: Solar Dynamics Observatory/Helioseismic and Magnetic Imager
\item
HELAS: European Helio- and Asteroseismology Network
\item
MHD: Magnetohydrodynamics
\item
MAG waves: Magneto-Acoustic-Gravity waves
\item
OLA: Optimally Localized Averaging (or Averages)
\item
RLS: Regularized Least Squares
\end{enumerate}

\section*{KEY TERMS/DEFINITIONS}
\begin{enumerate}
\item
Active region: Region of enhanced magnetic activity, including sunspots and diffuse magnetic field (`plage').
\item
Quiet Sun: Regions with low levels of magnetic activity, away from active regions.
\item
Dopplergram: Image of the line-of-sight component of velocity of the solar surface.
\item
The forward problem: The problem of computing the propagation of waves through a given solar model.
\item
The inverse problem: The problem of inferring solar subsurface properties from helioseismology measurements.
%\item
%full-waveform forward modeling: Constructing a model of the Sun that is consistent with
%the observations of solar oscillations.
\item
Ring-diagram analysis: Analysis of the local frequencies of solar oscillations
over small patches of the solar disk.
\item
Cross-covariance: Measure of similarity of two random signals as a function of a time-lag applied to one of them.
\item
Time-distance diagram: cross-covariance of the helioseismic signal between two points on the surface, as a function of their separation distance and time lag.
\item
Farside: Side of the Sun that is not visible from the Earth.
\end{enumerate}

\section*{ANNOTATED REFERENCES}
\begin{enumerate}
\item  \citet{Bogdan1997}: Solar modes, wave packets, and rays.
\item  \citet{Braun1995}: Mode absorption and mode coupling by sunspots.
\item  \citet{Cameron2008}:  Observations and modeling of the cross-covariance around a sunspot.
\item  \citet{Giles1997}: Inferring meridional circulation with time-distance helioseismology.
\item  \citet{Gizon2002}: The forward problem and the first Born approximation.
\item  \citet{Gizon2005}: Comprehensive open-access review of local helioseismology.
\item  \citet{Jefferies2006}: Multi-height observations of solar oscillations and magnetic portals.
\item  \citet{Komm2004}: Ring-diagram analysis of subsurface flows.
\item  \citet{Kosovichev2000}: Review of time-distance helioseismology.
\item  \citet{Lindsey2000SCI}: Imaging active regions on the farside of the Sun.
\end{enumerate}

\section*{RELATED RESOURCES}
\begin{enumerate}
%\item    Living Reviews of Solar Physics, ``Local Helioseismology'' by L. Gizon and A.C. Birch. Open access article at \url{http://livingreviews.org/lrsp-2005-6}
\item
Instrument web sites: GONG web site at \url{http://gong.nso.edu/} and SOHO/MDI at \url{http://soi.stanford.edu/}.
\item
MDI Farside Graphics Viewer at \url{http://soi.stanford.edu/data/full_farside/farside.html}.
\item
HELAS local helioseismology web site at \url{http://www.mps.mpg.de/projects/seismo/NA4/}. Software tools and selected data sets.
\item
Solar Physics, Vol. 192, No. 1-2, pp. 1-494 (2000), Topical Issue ``Helioseismic Diagnostics of Solar Convection and Activity'' edited by T.L. Duvall Jr., J.W. Harvey, A.G. Kosovichev, and Z. Svestka. Table of contents available at
\url{http://www.springerlink.com/content/h4bhbw3vdj8n/}.
%\url{www.iop.org/EJ/toc/1742-6596/118/1}
\item
Solar Physics, Vol. 251, No. 1-2, pp. 1-666 (2008), Topical Issue ``Helioseismology, Asteroseismology, and MHD Connections'' edited by L. Gizon, P. Cally, and J. Leibacher. Table of contents available at
\url{http://www.springerlink.com/content/x548678p1725/}.
\end{enumerate}

\section*{SIDE BAR: Extracting information from a random wave field}
\citet{Duvall1993} first used the cross-covariance function to measure the travel time of wave packets between two locations on the solar surface. The cross-covariance averages the information over an ensemble of random waves, constructively. The concept of time-distance helioseismology has found many applications in physics, geophysics, and ocean acoustics \citep[see reviews by][]{Larose2006,Gouedard2008}. Various experiments and observations \citep[e.g.][]{Weaver2001, Shapiro2005} have shown that the cross-covariance is intimately connected to the Green's function, $G$, i.e. the response of the medium to an impulsive source. Recently, \citet{Colin2006} proved that in an {\it arbitrarily complex} medium containing an homogeneous distribution of white noise sources (variance $\sigma^2$), the cross-covariance is given by
\begin{equation}
\frac{\partial}{\partial t}C(\br_{1},\br_{2}, t)
= -\frac{\sigma^2}{4 a}\left[G(\br_{1}, \br_{2},t) + G(\br_{2}, \br_{1}, -t) \right] ,
\end{equation}
when the integration time tends to infinity and the coefficient of attenuation ($a$) tends to zero. In the Fourier domain, this is equivalent to saying that $C$ is proportional to the imaginary part of the Green's function, ${\rm Im}\; G(\br_{1}, \br_{2}, \omega)$. Although the above assumptions are too restrictive to be applied to the solar case, it is clear that the cross-covariance is a very important diagnostics to probe media permeated by random fields (wave fields or diffuse fields).

\section*{DISCLOSURE STATEMENT}
The authors are not aware of any biases that might be perceived as affecting
the objectivity of this review.

\section*{ACKNOWLEDGEMENTS}
LG acknowledges support from the Deutsches Zentrum
f{\"u}r Luft- und Raumfahrt through project `German Data Center for the Solar
Dynamics Observatory,' from the European Research Council through FP-7
Starting Grant `Seismic Imaging of the Solar Interior', and from the European
Union through FP-6 Coordination Action `European Helio- and
Asteroseismology Network'. ACB acknowledges support from NASA contracts NNH06CD84C,
NNH07CD25C, NNG07EI51C, and NNH09CE41C.
SOHO is a mission of international collaboration between  ESA and NASA. GONG is an
international collaboration sponsored by the NSF
through a cooperative agreement with the Association of Universities for
Research in Astronomy Inc. The MDI and HMI projects are supported by NASA under contracts to Stanford
University. We are very grateful to Thomas Bogdan, Doug Braun, Paul Cally, Robert
Cameron, Ashley Crouch, Alina Donea, Thomas Duvall Jr., Bernhard Fleck, Irene Gonz\'alez
Hern\'andez, Deborah Haber, Frank Hill, Jason Jackiewicz, Rudi Komm, Alexander Kosovichev, John Leibacher, Charles Lindsey, Hamed Moradi,
Matthias Rempel,
Philip Scherrer, and Hannah Schunker for providing figures and/or comments.

\section*{SUPPLEMENTAL MATERIAL}
11 Movies available online.

%\section*{APPENDIX}
%We may need, later, an additional appendix to cover some of the results of the SDO mission, expected to be launched in February 2010.

\end{document}